\documentclass[preprint,aps,showpacs,nofootinbib,preprintnumbers,amsmath,amssymb]{revtex4-1}
\usepackage{amssymb}
\usepackage{epsfig}
\usepackage{graphicx}
\usepackage{subfigure}
\usepackage{dcolumn}
\usepackage{bm}
\usepackage[usenames ,dvipsnames]{xcolor}
\usepackage{slashed}

\begin{document}

\title{Strongly self-interacting vector dark matter via freeze-in}

\author{Mateusz Duch\footnote{mateusz.duch@fuw.edu.pl}, Bohdan Grzadkowski\footnote{bohdan.grzadkowski@fuw.edu.pl}, and
Da~Huang\footnote{da.huang@fuw.edu.pl}
}
 \affiliation{Faculty of Physics, University of Warsaw, Pasteura 5, 02-093 Warsaw, Poland}

\date{\today}
\begin{abstract}
We study a vector dark matter (VDM) model in which the dark sector couples to the Standard Model sector via a Higgs portal. If the portal coupling is small enough the VDM can be produced via the freeze-in mechanism. It turns out that the electroweak phase transition have a substantial impact on the prediction of the VDM relic density. We further assume that the dark Higgs boson which gives the VDM mass is so light that it can induce strong VDM self-interactions and solve the small-scale structure problems of the Universe. As illustrated by the latest LUX data, the extreme smallness of the Higgs portal coupling required by the freeze-in mechanism implies that the dark matter direct detection bounds are easily satisfied. However, the model is well constrained by the indirect detections of VDM from BBN, CMB, AMS-02, and diffuse $\gamma$/X-rays. Consequently, only when the dark Higgs boson mass is at most of ${\cal O}({\rm keV})$ does there exist a parameter region which leads to a right amount of VDM relic abundance and an appropriate VDM self-scattering while satisfying all other constraints simultaneously.
\end{abstract}

\maketitle

\section{Introduction}
\label{s1}
In spite of increasing astrophysical and cosmological evidence for the existence of the dark matter (DM)~\cite{PDG,Bergstrom:2012fi}, the nature of DM remains a mystery. According to the dominant paradigm DM consists of collisionless, cold particles that successfully explain the large scale structures in our Universe. However, collisionless cold DM predictions obtained by N-body simulations face some difficulties known e.g. as the cusp-vs-core problem~\cite{Moore:1994yx,Flores:1994gz,Oh:2010ea,Walker:2011zu} or the too-big-to-fail problem~\cite{BoylanKolchin:2011de,BoylanKolchin:2011dk,Garrison-Kimmel:2014vqa} when confronted with precise observations at the dwarf scale. However, it has been shown that the presence of sizable DM self-interactions with $\sigma_{\rm DM}/m_{\rm DM} = 0.1 \sim 10~{\rm cm^2/g}$ has the potential to alleviate such a tension~\cite{deLaix:1995vi,Spergel:1999mh,Vogelsberger:2012ku,Zavala:2012us,Rocha:2012jg, Peter:2012jh,Kaplinghat:2015aga,Tulin:2017ara}, even though the DM self-interactions are constrained to be  $\sigma_{\rm DM}\lesssim 1~{\rm cm^2/g}$ by measurements at the cluster scale~\cite{Clowe:2003tk,Markevitch:2003at,Randall:2007ph,Kahlhoefer:2013dca,Harvey:2015hha,Wittman:2017gxn}.

Such large DM self-scatterings naturally arise if there is a light particle mediating the DM interaction and the corresponding cross-section is enhanced by non-perturbative effects~\cite{Buckley:2009in,Loeb:2010gj,Ackerman:mha,Feng:2009hw,Tulin:2012wi,Tulin:2013teo, Aarssen:2012fx,Cyr-Racine:2015ihg,Nozzoli:2016coi,Foot:2014uba}. One immediate consequence of this light mediator scenario is that the DM self-interaction cross section is velocity dependent~\cite{Markevitch:2003at,Randall:2007ph, Peter:2012jh,Rocha:2012jg, Kahlhoefer:2013dca,Harvey:2015hha, Kaplinghat:2015aga}, which allows for the signals at the dwarf scale to evade the constraints from the galaxy clusters. A simple way to realize this scenario is to introduce a model, where DM is generated via the dark freeze-out mechanism in which it predominantly annihilates into a pair of light mediators. Nevertheless, it has recently been shown in Refs.~\cite{Bringmann:2016din,Kahlhoefer:2017umn} that this secluded DM model~\cite{Pospelov:2007mp} is severely constrained by the DM indirect detection. A way to avoid these problems is to consider a DM production mechanisms different from the conventional freeze-out. One possibility is the freeze-in mechanism~\cite{McDonald:2001vt,Hall:2009bx} (see {\it i.g.} Ref.~\cite{Bernal:2017kxu} for a recent review and the complete references therein). It is found in Refs.~\cite{McDonald:2001vt,Hall:2009bx,Bernal:2017kxu,Cheung:2010gj,Chu:2011be} that the final DM relic density is determined exclusively by the main DM production channels at the freeze-in temperature and it is not sensitive to many details of DM evolution at higher temperature, which guarantees the predictability of this mechanism.

The freeze-in as a production mechanism for self-interacting dark matter was analyzed in \cite{Campbell:2015fra,Kang:2015aqa,Bernal:2015ova,Bernal:2015xba,Ayazi:2015jij,Bernal:2017mqb}. Notably, the case of light-mediator was discussed in ref.~\cite{Bernal:2015ova} within the model of Hidden Vector DM with dark $SU(2)$ gauge symmetry \cite{Hambye:2008bq}, where it has been found that the scenario with keV mediator agrees with experimental constraints. It has been also noticed that if decays of the mediator into $e^+ e^-$ are allowed, its significant abundance and large lifetime cannot satisfy bounds from Big Bang Nucleosynthesis (BBN), so that this region of the parameters is excluded.

In this work, we study an abelian {version of vector dark matter (VDM) models}~\cite{Hambye:2008bq,Lebedev:2011iq,Farzan:2012hh,Baek:2012se,Baek:2014jga, Duch:2015jta,Duch:2015cxa,Karam:2015jta,Karam:2016rsz,Heikinheimo:2017ofk, Arcadi:2016kmk} in which the VDM particle with mass of ${\cal O}({\rm GeV \sim TeV})$ couples to the SM sector only through the Higgs portal. We take into account recent bounds from BBN, CMB and discuss possibility of constraining the model with FERMI-LAT, AMS-02, diffuse $\gamma$/$X$-Ray and direct detection LUX data. In the case of indirect constraints on DM annihilation, we include the effect of Sommerfeld enhancement. We also take into account consequences of electro-weak phase transition in calculation of DM production. 
The dark Higgs boson of the VDM model is assumed to be so light that it can induce large self-interactions to solve the small-scale structure problems. We focus on the scenario in which the VDM is produced by the freeze-in mechanism. The main question that we address is if there exist a region in the parameter space that can generate the right VDM relic abundance and appropriate DM self-scatterings while still satisfying all the direct and indirect detection constraints. After scanning over the parameter space we conclude that if the mediator $h_2$ is too light to decay into $e^+ e^-$, then indeed all the constraints can be satisfied together with correct relic abundance and appropriate DM self-scatterings. The necessary mediator mass is of the order of ${\cal O}({\rm keV})$.
Our results agree with those found in \cite{Bernal:2015ova}.

The paper is organized as follows. In Sec.~\ref{Sec_Model}, the VDM model is presented. The VDM production via freeze-in is discussed in Sec.~\ref{Sec_FI}, with a special attention to the effects of the electroweak (EW) phase transition. Then we discuss the VDM self-interactions in Sec.~\ref{SecSI}. Sec.~\ref{Sec_DD} and Sec.~\ref{Sec_ID} are devoted to constraints from DM direct and indirect detection experiments. The numerical results are presented in Sec.~\ref{Sec_Res}. Finally, we give a brief summary in Sec.~\ref{Sec_Conc}. Some useful formulae are collected in Appendix~\ref{appA}.

\section{The Model}\label{Sec_Model}
Following Refs.~\cite{Duch:2015jta,Duch:2015cxa}, we introduce a dark $U(1)_X$ gauge symmetry and a complex scalar $S$ which is neutral under SM gauge group but has unit charge under this $U(1)_X$ symmetry. We further assume an additional $Z_2$ symmetry, under which the gauge boson $X_\mu$ and $S$ transform as follows:
\begin{eqnarray}
X_\mu \to -X_\mu\,,\;\;\;S\to S^*\,,
\label{cc}
\end{eqnarray}
which is just the charge conjugate symmetry in the dark sector. It forbids the kinetic mixing between the SM $U(1)_Y$ gauge boson $B_\mu$ and $X_\mu$,  $X_{\mu\nu} B^{\mu\nu}$, ensuring stability of $X_\mu$. Therefore, the relevant dark sector Lagrangian is given by
\begin{eqnarray}
{\cal L}_{\rm d} = -\frac{1}{4}X_{\mu\nu}X^{\mu\nu} + (D_\mu S)^\dagger D^\mu S {+ \mu_S^2 |S|^2 - \lambda_S|S|^4 - \kappa |S|^2 |H|^2},
\end{eqnarray}
where $H$ is the usual SM Higgs $SU(2)_L$ doublet, and the covariant derivative of $S$ is defined as $D_\mu S \equiv (\partial_\mu+i g_X X_\mu)S$ with $g_X$ being the corresponding dark gauge coupling constant. Note that the quartic portal interaction, $\kappa |S|^2 |H|^2$, is the only connection between the dark sector and the SM, so in the limit $\kappa\to 0$ the two sectors decouple.
Also, the mass term of $S$ has the negative sign compared with the usual scalar field, so that it can induce the spontaneous symmetry breaking (SSB) of the gauge $U(1)_X$. By minimizing the scalar potential of the model, we can obtain the vacuum expectation values of the usual SM Higgs doublet $\langle H \rangle \equiv (0, v_H/\sqrt{2})^T$ and the dark Higgs $\langle S \rangle \equiv v_S/\sqrt{2} $ as follows:
\begin{eqnarray}
v_H^2 = \frac{4 \lambda_S \mu^2_h - 2\kappa \mu_S^2}{4 \lambda_H \lambda_S -\kappa^2} \, ,\quad\quad v_S^2 =  \frac{4 \lambda_H\mu_S^2 - 2\kappa \mu_H^2}{4\lambda_H \lambda_S -\kappa^2}\,.
\end{eqnarray}
Note that $\langle S \rangle$ can be always assumed real without compromising any generality, therefore the discrete symmetry (\ref{cc}) remains unbroken as needed for the stability of $X_\mu$.

After the SSB happens, the dark gauge boson obtains its mass $m_X = g_X v_S$ via the dark Higgs kinetic term, and both scalar fields can be written as
\begin{eqnarray}\label{ScalarBEW}
H=\left( \begin{array}{c}
           H^+ \\
           (v_H + \phi_H +i\sigma_H)/\sqrt{2}
         \end{array}
\right)\,,\quad\quad S= \frac{1}{\sqrt{2}} (v_S + \phi_S +i\sigma_S)\,.
\end{eqnarray}
By expanding the scalar potential up to the second order, the mass squared matrix ${\cal M}^2$ of the two physical scalars $(\phi_H,\phi_S)^T$ is given by
\begin{eqnarray}
{\cal M}^2 = \left( \begin{array}{cc}
                      2\lambda_H v_H^2 & \kappa v_H v_S \\
                      \kappa v_H v_S & 2\lambda_S v_S^2
                    \end{array}
 \right)\,.
\end{eqnarray}
With the following orthogonal transformation of scalars,
\begin{eqnarray}\label{ScalarTrans}
\left( \begin{array}{c}
         \phi_H \\
         \phi_S
       \end{array}
 \right) = \left( \begin{array}{cc}
                    c_\theta & -s_\theta \\
                    s_\theta & c_\theta
                  \end{array}
  \right) \left( \begin{array}{c} h_1 \\ h_2 \end{array} \right)\,
\end{eqnarray}
we can define the mass eigenstates $(h_1, h_2)^T$ with their masses $(m_{h_1}, m_{h_2})$, where $\theta$ is the mixing angle with $s_\theta \equiv \sin \theta$ and $c_\theta \equiv \cos\theta$. As a result, we have the following relations:
\begin{eqnarray}\label{def_k}
\kappa = \frac{(m_{h_1}^2 - m_{h_2}^2)s_{2\theta}}{2 v_H v_S}\,,\quad {\lambda_H = \frac{m_{h_1}^2 c_\theta^2 + m_{h_2}^2 s_\theta^2}{2 v_H^2}\, ,\quad \lambda_S = \frac{m_{h_2}^2 c_\theta^2 + m_{h_1}^2 s_\theta^2}{2v_S^2}}\,.
\end{eqnarray}
In the freeze-in mechanism, the dark sector composed of $X$ and $h_2$ never thermalizes with the visible SM sector, so that the portal interactions $\kappa$ or $s_\theta$ should be very tiny. As is evident from Eq.~(\ref{ScalarTrans}), the $h_1$ boson is mostly SM-Higgs-like, while $h_2$ is almost the dark Higgs $\phi_S$. We have found that the most convenient choice of input parameters which specify the models is ($m_X, \,m_{h_2},\, \kappa,\, g_X $), together with the already known parameters $v_H = 246$~GeV and $m_{h_1} = 125$~GeV.

\section{Vector Dark Matter Relic Density via Freeze-In}\label{Sec_FI}
Within the freeze-in mechanism, the standard assumption is that the initial abundances of the VDM and the dark Higgs $h_2$ after reheating are assumed to be negligibly small, which is possibly a result of the reheating process itself or another mechanism. Furthermore, the Higgs portal coupling should be very tiny so that the dark sector can neither thermalize itself nor be in equilibrium with the SM sector. When the VDM mass is smaller than the EW phase transition temperature $T_{\rm EW} \simeq 160$~GeV~\cite{Quiros:1999jp,Katz:2014bha} its abundance is mainly controlled by various SM particle annihilations and/or decays that contribute to the collision term of the following Boltzmann equation
\begin{eqnarray}\label{BEq}
xHs\frac{dY_X}{dx} = \sum_f \gamma_f + \gamma_W +\gamma_{h_1}+\gamma_Z + \gamma^D_{h_1}\,,
\end{eqnarray}
where $Y_X = n_X/s$ is the DM yield defined as a ratio of DM number density $n_X$ and  the entropy density in the visible sector $s$. The parameter $x\equiv m_X/T$ describes the SM sector temperature $T$, $H$ is the Hubble parameter, and $\gamma_i \equiv \langle \sigma v \rangle_i (n^i_{{\rm eq}})^2$ denotes the so-called reaction density~\cite{Chu:2011be} for the SM particles annihilation into VDMs (for $\gamma_f$ we sum over all SM fermions).
The last term
represents the SM-Higgs-like $h_1$ decays to a VDM pair when this channel is kinematically allowed.
Since here $m_{h_2}\ll m_X$, no corresponding decay term for $h_2$ appears.
In this project, the model is implemented within \texttt{LanHEP}~\cite{LanHEP1,LanHEP3} and calculations of the cross sections and decay rates are performed adopting \texttt{CalcHEP}~\cite{CalcHEP}. Definitions of reaction densities, obtained cross sections and decay rates are collected in Appendix~\ref{appA}.

\begin{figure}[ht]
\includegraphics[scale = 0.7]{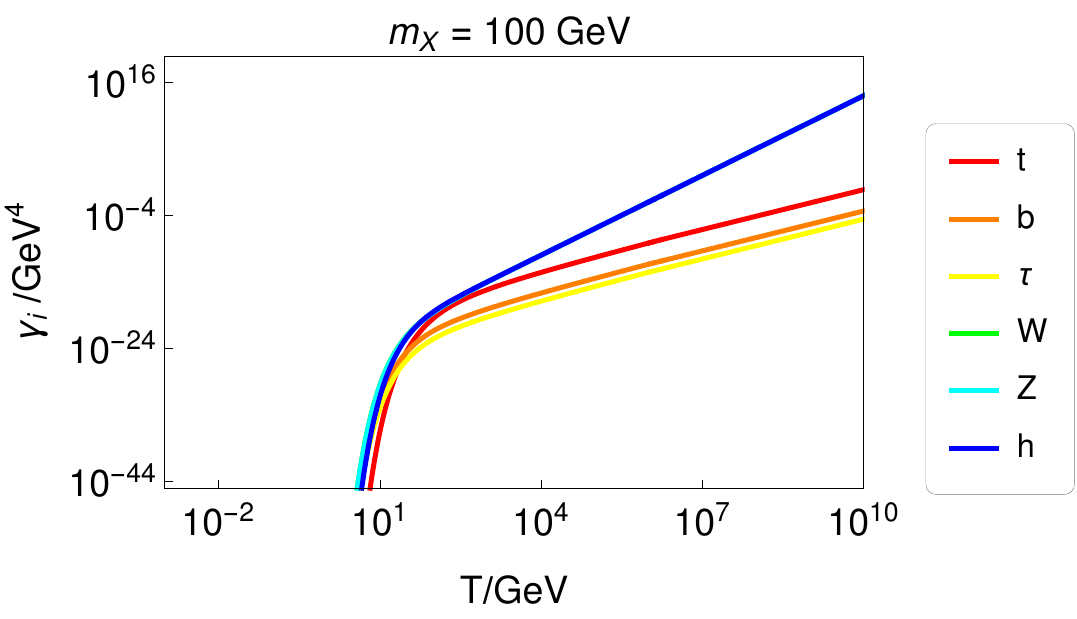}
\includegraphics[scale=0.7]{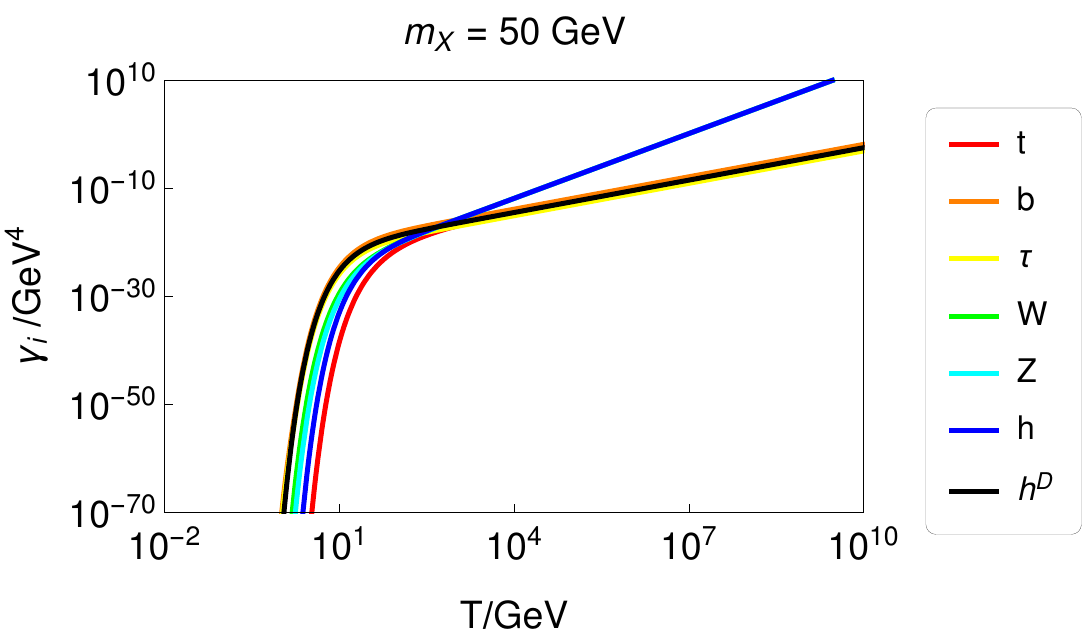}
\caption{Evolution of the reaction densities for various SM channels as functions of the SM sector temperature $T$ for $m_X = 100$~GeV and $\kappa=2.09 \times 10^{-11}$ (left panel) and $m_X = 50$~GeV and $\kappa = 2.40\times 10^{-12}$ (right panel). The chosen values of $\kappa$ result in the observed DM relic abundance.
}\label{Rates}
\end{figure}

\begin{figure}[ht]
\includegraphics[scale = 0.6]{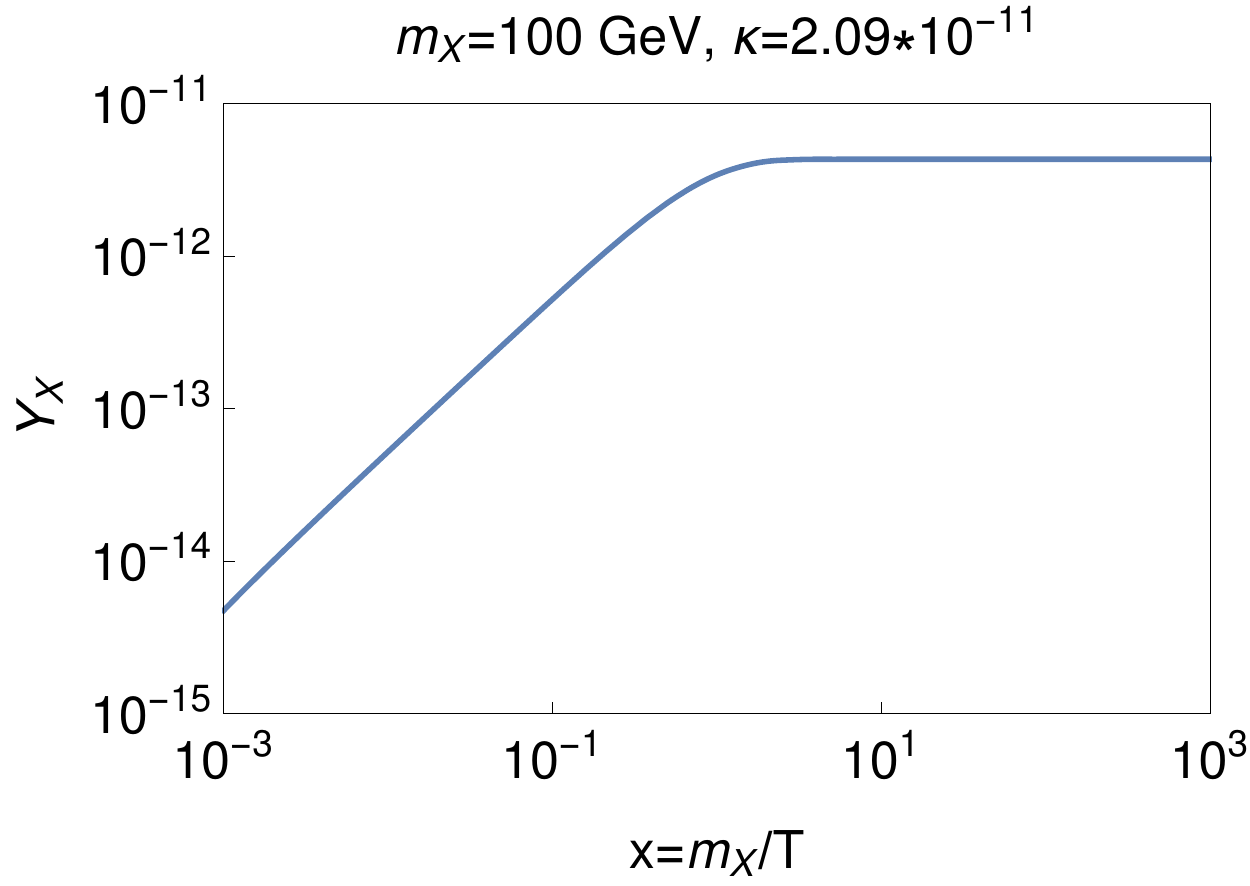}
\includegraphics[scale=0.6]{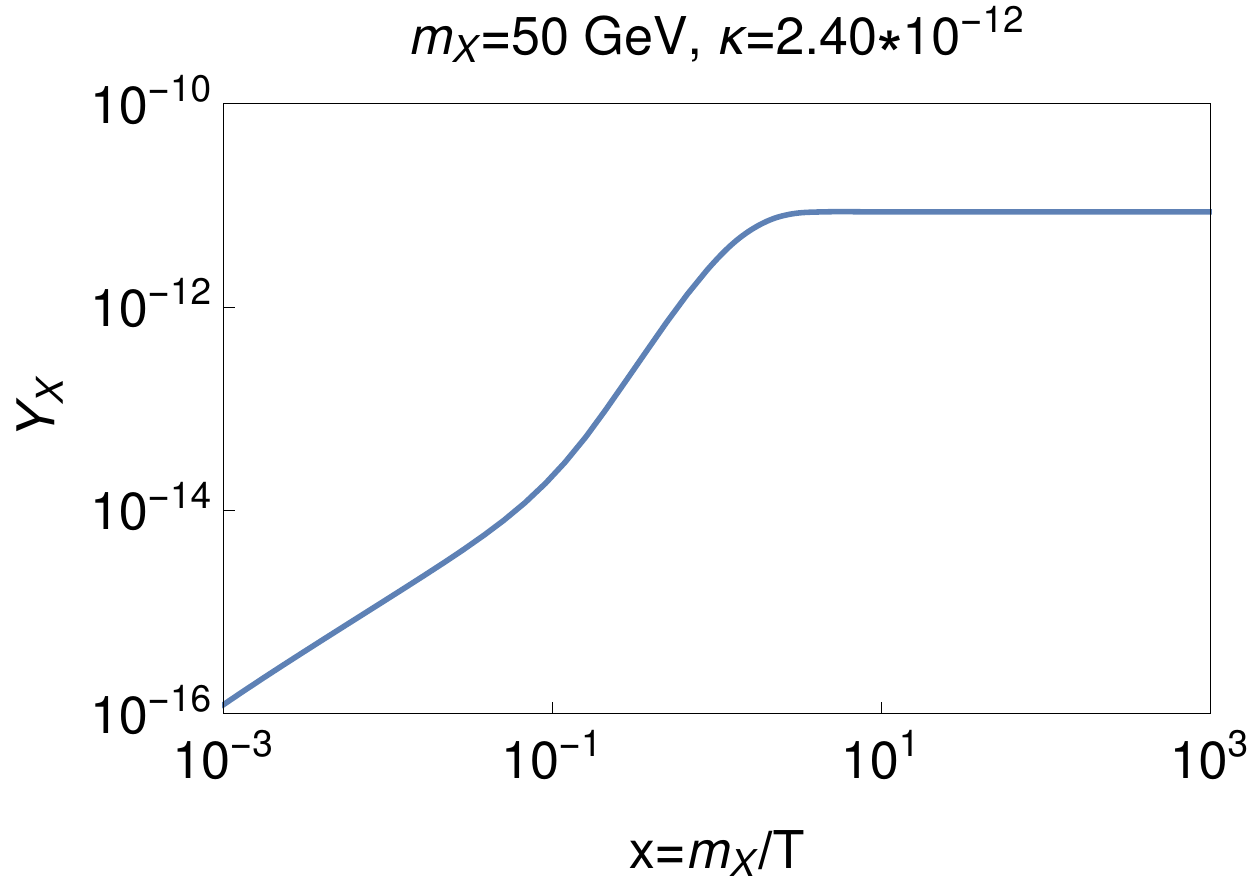}
\caption{Examples of evolutions of the VDM yield $Y_X$ as the functions of $x=m_X/T$, in which the model parameters for both panels are chosen the same as those in Fig.~\ref{Rates}.}\label{Yields}
\end{figure}

It is interesting to note that all of the reaction densities are proportional to the square of the Higgs portal coupling $\kappa^2$ with no dependance on $g_X$~\footnote{This property could be easily seen adopting the first relation of (\ref{def_k}).}, which explains why we have decided to use $\kappa$ instead of {$\sin\theta$} as a parameter. Also, due to the assumed mass hierarchy $m_{h_2}\ll m_X$, the value of $m_{h_2}$ influences the resulting DM abundance very weakly. Hence, the prediction for VDM relic abundance depends mainly on two parameters $m_{X}$ and $\kappa$. Fig.~\ref{Rates} shows typical examples of evolution of reaction densities for various SM channels as functions of the SM sector temperature $T$. The left and right panels represent the case with $m_X$ larger or smaller than $m_{h_1}/2$. It is evident that in the first case, only the annihilations of the SM particles are involved. On the other hand in the second scenario the decay of the visible Higgs $h_1$ is open and overwhelms other annihilation modes near the freeze-in temperature $T_{\rm FI}$~\footnote{Defined as the temperature at which the VDM yield from freeze-in production reaches its maximal value and stabilizes.}. Since the freeze-in mechanism is IR dominated~\cite{Hall:2009bx,Chu:2011be}, the VDM relic density is dictated by the $h_1 \to X X$ decay rate. We present the resulting evolution of the VDM yields $Y_X$ in Fig.~\ref{Yields}, which illustrate typical features of the freeze-in mechanism.

\begin{figure}[ht]
\includegraphics[scale = 0.8]{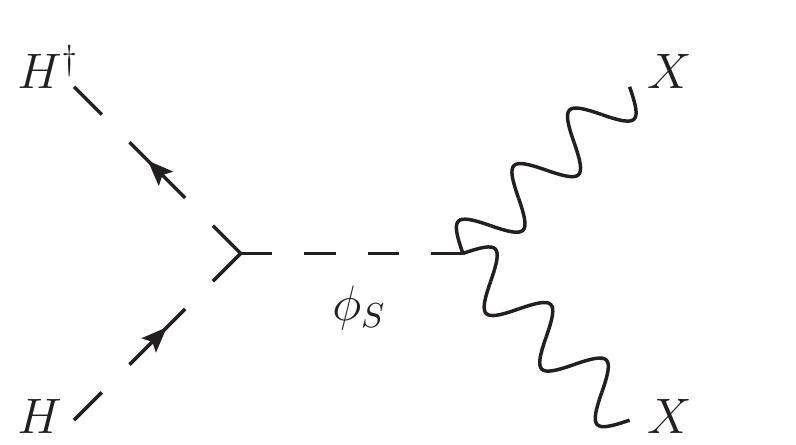}
\caption{Feynman diagram to generate the VDM $X$ via the SM Higgs doublet $H$ annihilations. }\label{HHbar}
\end{figure}

However, when the VDM mass is much larger than the EW phase transition temperature $T_{\rm EW}$, the VDM abundance stops increasing before the EW phase transition. In this case, the SM gauge symmetry $SU(2)_L \times U(1)_Y$ is not broken, so that only the tree-level diagram shown in Fig.~\ref{HHbar} can generate VDM particles. Hence, the Boltzmann equation can be simplified to:
\begin{eqnarray}\label{BEqB}
xHs\frac{dY_X}{dx} = \gamma_{H\bar{H}}\,.
\end{eqnarray}

\begin{figure}[ht]
\includegraphics[scale = 0.7]{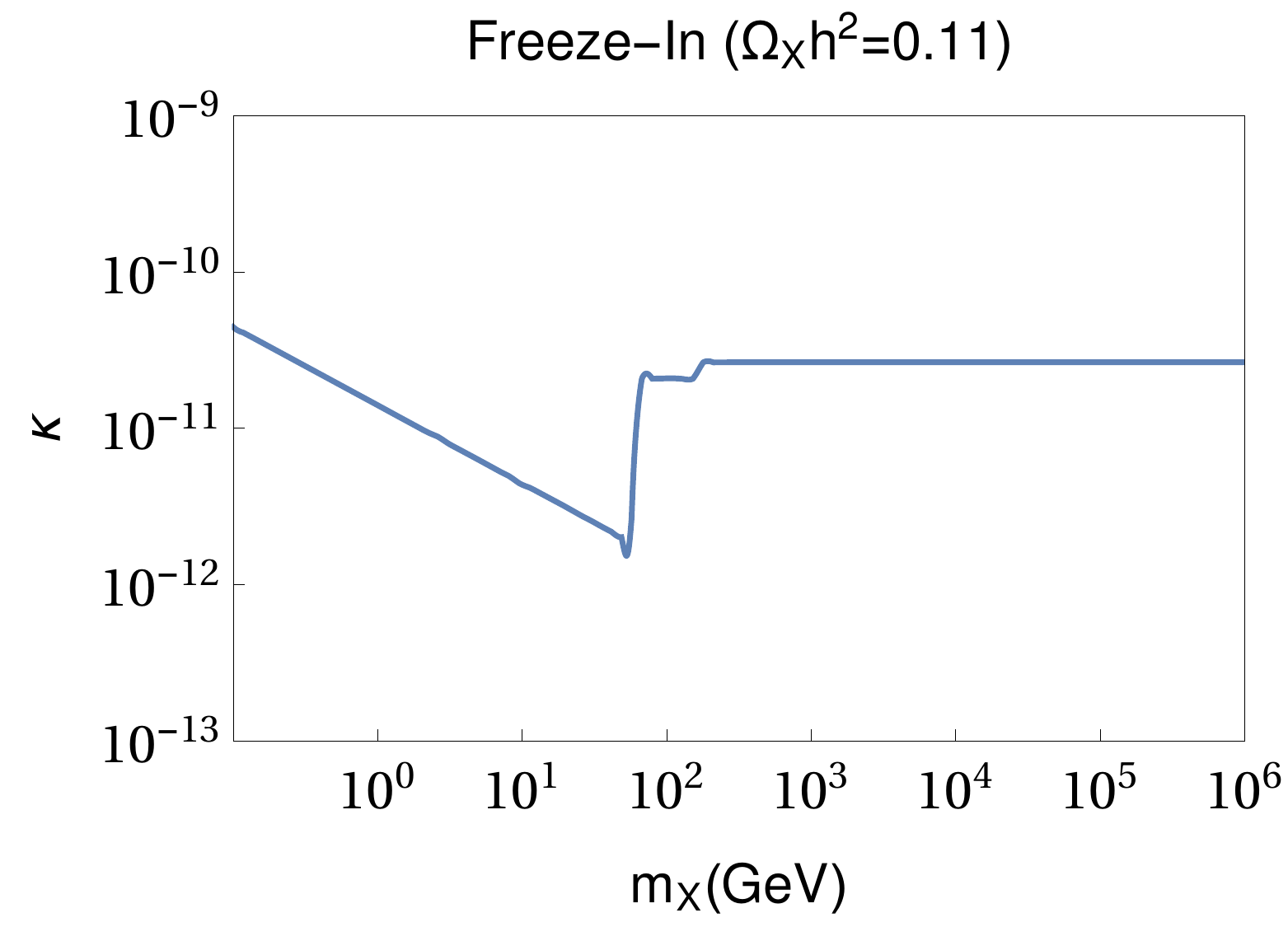}
\caption{The {value} of $\kappa$ as a function of the VDM mass $m_X$, that gives the observed relic density via the freeze-in mechanism.}\label{RelicDensity}
\end{figure}

By comparing the solutions to the Boltzmann equations in Eqs.~(\ref{BEq}) and (\ref{BEqB}) with the observed DM relic density $\Omega_{\rm X}h^2 =0.11$, we can obtain the value of $\kappa$ as the function of the VDM mass $m_X$ in Fig.~\ref{RelicDensity}.
It is interesting to see the change of $\kappa$-$m_X$ scaling in this plot, which reflects the transitions of the dominant VDM freeze-in channels. When the VDM mass is larger than $m_{h_1}/2$, as mentioned before, only the annihilation modes contribute, no matter whether the EW gauge symmetry is broken or not. In this case, according to the argument in Ref.~\cite{Hall:2009bx,Cheung:2010gj}, the yield could be estimated as
\begin{eqnarray}\label{rela1}
Y_X \sim \sigma (T_{\rm FI}) M_{\rm Pl} T_{\rm FI} \sim \kappa^2 \frac{M_{\rm Pl}}{T_{\rm FI}} \sim \kappa^2 \frac{M_{\rm Pl}}{m_X}\,,
\end{eqnarray}
where the first relation follows from the dimensional argument with $M_{\rm Pl}$ being the Planck mass. $\sigma(T_{\rm FI})$ is the total cross section of the SM particle annihilation at the freeze-in temperature $T_{\rm FI}$, which is simplified to be $\sigma \sim \kappa^2/T^2_{\rm FI}$.
We have also used the relation $T_{\rm FI} \sim m_X$, which can be understood as follows. When $m_X > T_{\rm EW}$, as it has been mentioned above only the channel $H H^\dagger \to XX$ contributes to VDM generation. It becomes ineffective as the temperature drops below $m_X$, since then the SM Higgs doublets do not have enough kinetic energies. On the other hand, for the case with $m_X \leq T_{\rm EW}$, the VDM freeze-in process is dominated by the annihilations of particles which are lighter than the VDM. Similarly, when the SM plasma temperature decreases below $m_X$, the VDM yield ceases to grow any more due to the fact that these channels are no longer kinematically allowed. Concluding, the freeze-in temperature is expected to be around the VDM mass, $T_{\rm FI}\sim m_X$, in the present scenario. Then it is easy to derive from Eq.~(\ref{rela1}) that the predicted VDM relic density $\Omega_X h^2 \propto Y_X m_X $ should only depend on $\kappa$ whereas the dependence on $m_X$ are cancelled out, which is manifested as a flat line in Fig.~\ref{RelicDensity}. However, if the VDM is lighter than a half of the visible Higgs mass, the decay channel $h_1 \to X X$ dominates, so that the VDM yield should be
\begin{eqnarray}
Y_X \sim \Gamma_{h_1 \to XX} \frac{M_{\rm Pl}}{T_{\rm FI}^2} \sim \kappa^2 m_{h_1} \frac{M_{\rm Pl}}{T_{\rm FI}^2} \sim \kappa^2 \frac{M_{\rm Pl}}{m_{h_1}}\,,
\end{eqnarray}
where the decay rate should be $\Gamma_{h_1 \to XX} \sim \kappa^2 m_{h_1}$, and the freeze-in temperature in this case is $T_{\rm FI} \sim m_{h_1}$ at which the density of the visible Higgs $h_1$ is greatly suppressed by its Boltzmann factor. Hence, the VDM relic density is $\Omega_X h^2 \propto \kappa^2 m_X$, which results in the scaling of $\kappa\propto m_X^{-1/2}$ in Fig.~\ref{RelicDensity}. Finally, note that the small but abrupt rise of $\kappa$ at the $m_X = 160$~GeV represents the EW phase transition effect due to the sudden change of the main VDM production channels.


{In order for the freeze-in mechanism to work, it is required that the dark sector neither thermalize by itself nor with the SM sector. It is easy to check that the portal coupling $\kappa$ implied by the VDM relic density is so tiny that it is impossible for the dark sector to equilibrate with the visible one. However, the non-thermalization of the dark sector  by itself is not guaranteed. When the number densities of the VDM and $h_2$ accumulated via freeze-in become large enough, it is probable that the dark sector process $XX \to h_2 h_2$ would be cosmologically efficient, which would soon change the number densities of VDM and $h_2$ to form a dark plasma with a common (and in general different from the SM)  temperature. Therefore, one should ensure that thermalization in the dark sector cannot take place and the appropriate condition can be coded by the following inequality~\cite{Chu:2011be,Bernal:2015ova}:
\begin{eqnarray}\label{RegionFI}
\langle \sigma(XX \to h_2 h_2)v\rangle n_X \leq H \,,
\end{eqnarray}
where $\langle \sigma(XX \to h_2 h_2)v\rangle$, $n_X$, and $H$ represent the thermally averaged cross section for VDM annihilations into $h_2$ pairs, the number density of VDM, and Hubble parameter, respectively, all of which are evaluated at the freeze-in temperature $T_{\rm FI}$. Note that $\langle \sigma(XX \to  h_2 h_2)v\rangle$ is proportional to the dark gauge coupling $\alpha_X^2$, so that it should not be suppressed in the parameter space where the DM has large self-interactions. Thus, the condition in Eq.~(\ref{RegionFI}) is not easy to be satisfied in the present scenario and therefore it constraints substantially the freeze-in parameter space as shown below. }

\section{Vector Dark Matter Self-Interactions via a Light Mediator}\label{SecSI}
It is well known that the cosmological small-scale structure problems, such as the 'cusp vs. core' and the 'too-big-to-fail' problems could be ameliorated if DM self-interaction was sufficiently strong at the dwarf galaxy scale~\cite{deLaix:1995vi,Spergel:1999mh,Vogelsberger:2012ku,Zavala:2012us,Rocha:2012jg, Peter:2012jh,Kaplinghat:2015aga,Tulin:2017ara}, the required value of the cross-section is
\begin{eqnarray}\label{DMSI}
0.1\,\frac{{\rm cm^2}}{\rm g} < \frac{\sigma_T}{m_X} < 10\, \frac{{\rm cm^2}}{\rm g} \,,
\end{eqnarray}
where $\sigma_T\equiv \int d\Omega (1-\cos \theta) d\sigma/d\Omega$ is the so-called momentum transfer cross section between DM particles. However,  DM self-scattering cross-section as large as $\sigma_T/m_X \simeq 10~{\rm cm^2/g}$ is not allowed by observations at the cluster scale with the typical constraint $\sigma_T/m_X < 1~{\rm cm^2/g}$~\cite{Clowe:2003tk,Markevitch:2003at,Randall:2007ph,Kahlhoefer:2013dca,Harvey:2015hha}.

A possible strategy that may generate large DM self-interaction is to introduce a mediator which is much lighter than the DM particles. In the VDM model, the elastic DM scattering is mediated by an exchange of the two Higgs scalars, $h_1$ and $h_2$. In the limit of small mixing, the $h_1$-mediated contribution is negligible due to $\sin\alpha$ and large $h_1$ mass suppression. In contrast, $XXh_2$ coupling is not suppressed by small mixing and, in addition, it is much lighter than the VDM particle, therefore $h_2$ can act as a light mediator which might be capable to amplify the self-interaction. When $\alpha_X m_X \ll m_{h_2}$ with $\alpha_X \equiv g_X^2/(4\pi)$ the fine-structure constant in the dark sector, the perturbative Born approximation is applicable in which the dominant $t$-channel $h_2$-exchange to the transfer cross section as follows~\cite{Tulin:2013teo}:
\begin{eqnarray}
\sigma_T^{\rm Born} =\frac{8\pi\alpha_X^2}{m_X^2 v^4} \left[\ln\left(1+ \frac{m_X^2 v^2}{m_{h_2}^2}\right)-\frac{m_X^2 v^2}{m_{h_2}^2 + m_X^2 v^2}\right]\,,
\end{eqnarray}
where $v$ is the relative velocity in the VDM two-body system.
Nevertheless, beyond the Born range, $h_2$ is much lighter than $\alpha_X m_X$ so that the nonperturbative effects would become important and give rise to the following attractive Yukawa potential:
\begin{eqnarray}\label{potential}
V(r) = -\frac{\alpha_X e^{-m_{h_2}r}}{r}\,.
\end{eqnarray}
Note that due to such nonperturbative corrections, the DM self-interactions have the non-trivial dependence on the VDM velocity. When the range of the potential characterized by $1/m_{h_2}$ is much larger than the VDM de Broglie wavelength $1/(m_X v)$, {\it i.e.}, $m_X v\gg m_{h_2}$, this part of parameter space is well known as the classical regime, for which analytic fitting formulas for $\sigma_T$~\cite{Khrapak:2003kjw,Tulin:2013teo,Tulin:2012wi,Cyr-Racine:2015ihg} are available in literature. In our numerical calculations, we adopt the more recent improved analytic expressions provided in Ref.~\cite{Cyr-Racine:2015ihg}. On the other hand, if $m_X v \lesssim m_{h_2}$, the VDM self-scatterings can be enhanced by several orders of magnitudes due to the formation of the quasi-bound states. This region of parameter space is usually denoted by the resonant regime. In this work, we obtain $\sigma_T$ in this regime by closely following Ref.~\cite{Tulin:2013teo} to solve the non-relativistic Schr$\ddot{\rm o}$dinger equation with the potential in (\ref{potential}). Moreover, it has been found~\cite{Khrapak:2003kjw,Tulin:2013teo,Tulin:2012wi,Cyr-Racine:2015ihg} that, with the presence of the non-perturbative effects, the VDM transfer cross section $\sigma_T$ is enhanced more significantly as the relative DM velocity becomes small. Such a velocity dependence of VDM self-scatterings is very appealing, since it helps the VDM model to solve the small-scale structure problems for the dwarf galaxy scale with a typical velocity $v\sim 10$~km/s while evading the strong constraints from the galaxy clusters with $v\sim 1000~{\rm km/s}$. More recently, a more careful analysis of DM self-interactions from a light (pseudo-)scalar has been presented in Ref.~\cite{Kahlhoefer:2017umn}, where a more appropriate definition of the momentum transfer cross section $\sigma_T$ is given and the possible correction from the $u$-channel light mediator exchange is investigated. However, it is seen in Ref.~\cite{Kahlhoefer:2017umn} that such corrections lead to very small modifications in final results so that we neglect them and follow the conventional formula from Refs.~\cite{Tulin:2013teo,Tulin:2012wi}.

\section{Direct Detection of the Vector Dark Matter}\label{Sec_DD}
It is usually claimed that the DM direct-detection experiments do not provide relevant constraints for models in which the DM particles are mainly produced by the freeze-in mechanism since the DM nuclear recoil cross sections are suppressed by tiny portal couplings. However, in the present scenario, the spin-independent (SI) VDM-nucleon (XN) scatterings are mediated by the two neutral Higgs bosons $h_{1,2}$, and thus it is possible that the cross-section is greatly enhanced by the small mass of the light mediator $h_2$. This feature is clearly reflected by the corresponding formula for the differential cross section of the XN scatterings with respect to the momentum transfer squared $q^2$,
\begin{eqnarray}\label{DDXS}
\frac{d\sigma_{XN}}{dq^2} = \frac{\sigma_{XN}}{4\mu_{XN}^2 v^2} G(q^2)\,,
\end{eqnarray}
where $v$ is the VDM velocity in the lab frame, $\mu_{XN} \equiv m_X m_N/(m_N+m_X)$ is the reduced mass of the XN system, and
\begin{eqnarray}
\sigma_{XN} = \frac{\kappa^2f_N^2 m_X^2 m_N^2 \mu_{XN}^2}{\pi m_{h_1}^4 m_{h_2}^2 (m_{h_2}^2 + 4 \mu_{XN}v^2)}
\end{eqnarray}
is the total cross section for the XN scattering with the effective nucleon coupling $f_N \approx 0.3$~\cite{Cline:2013gha,Alarcon:2011zs,Alarcon:2012nr}. Compared with the usual definition of the SI independent DM-nucleon cross section in the literature, Eq.~(\ref{DDXS}) has an additional form factor $G(q^2)$ defined as
\begin{eqnarray}\label{DefG}
G(q^2) = \frac{m_{h_2}^2 (m_{h_2}^2 + 4 \mu_{XN}^2 v^2)}{(q^2+m_{h_2}^2)^2}\,,
\end{eqnarray}
which encodes the effects of the light mediator $h_2$. It is clear that, for the heavy mediator case with $m_{h_2}^2 \gg q^2 \sim 4 \mu_{XN}^2 v^2$, the factor $G(q^2)$ will be reduced to 1, {\it i.e.}, we will recover the conventional XN contact interaction, and the usual experimental constraints can be applied. But when $m_{h_2}^2 \ll q^2$, the XN differential cross section in Eq.~(\ref{DDXS}) will have extra $q^2$ dependence characterized by the $G(q^2)$, thus modifying the corresponding nuclear recoil spectrum and, in turn, the final fitting results. Therefore, we need to re-analyze the experimental constraints in the latter case.

\begin{figure}[ht]
\includegraphics[width=.49\textwidth]{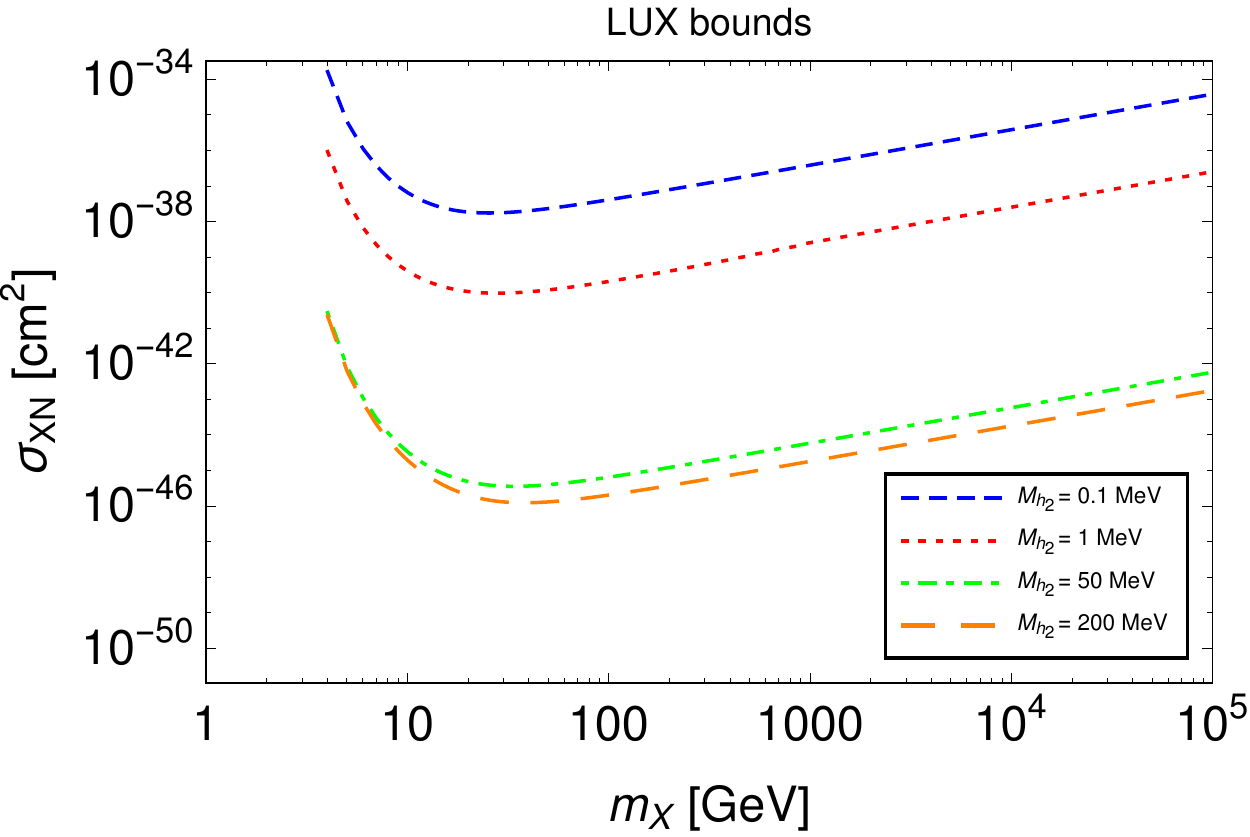}
\includegraphics[width=.49\textwidth]{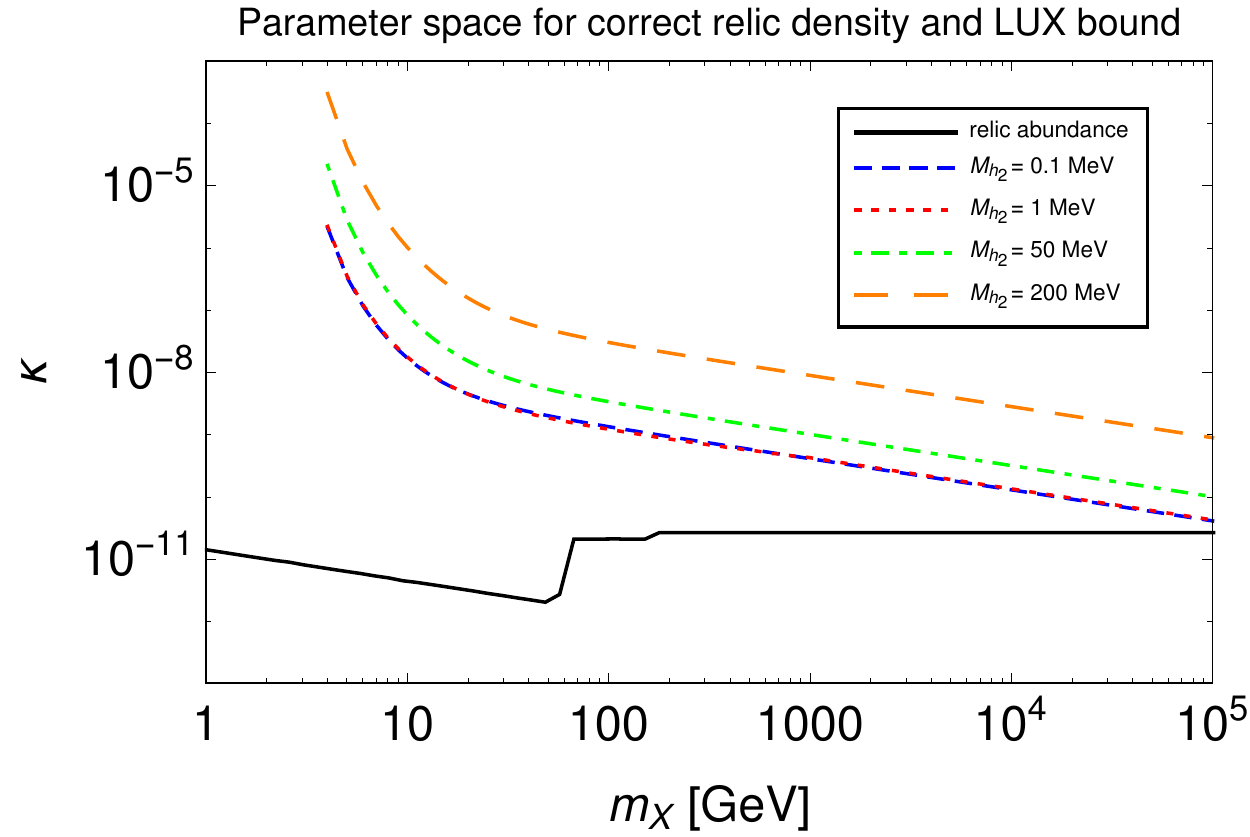}
\caption{LUX upper bounds on the total SI VDM nuclear recoil cross section $\sigma_{XN}$ (left panel) and Higgs portal coupling $\kappa$ (right panel) for different light mediator $h_2$ masses. The solid black curve in the right panel corresponds to parameters which reproduce the measured value of the DM relic density.} \label{Fig_DMDD}
\end{figure}

The strongest constraints on the direct detection of the VDM come from the LUX~\cite{Akerib:2016vxi}, PandaX-II~\cite{PandaX16} and XENON1T~\cite{XENON1T}. In the present work, we use the LUX 2016 dataset as an illustration of the SI direct detection limits to the VDM model since PandaX-II and XENON1T datasets would give the similar results. Due to the modification of the DM nuclear recoil spectrum caused by the light mediator $h_2$, we follow the simplified analysis methods presented in Ref.~\cite{Geng:2016uqt,Geng:2017ypy} to give the LUX 90\% C.L. upper bounds on the VDM nuclear scattering cross section $\sigma_{XN}$ and on the Higgs portal coupling $\kappa$, with the final numerical results as shown in Fig.~\ref{Fig_DMDD}.

It is seen from Fig.~\ref{Fig_DMDD} that the LUX upper bound on the VDM nuclear scattering cross section increases with the decrease of the the mediator $h_2$ mass. The lowest curve with $m_{h_2} = 200~{\rm MeV}$ corresponds to the point-like contact VDM nuclear interaction, and agrees with the upper limit in Ref.~\cite{Akerib:2016vxi}, since such a mass of $h_2$ is already much larger than the typical momentum transfer scale $q\sim 10$~MeV. However, when transformed into constraints on the Higgs portal coupling $\kappa$ on the right panel of Fig.~\ref{Fig_DMDD}, the LUX upper bound is found to behave oppositely, that is, it becomes stronger with the the smaller $m_{h_2}$. Furthermore, when $m_{h_2}\lesssim 1$~MeV, the LUX bound is shown to saturate a limiting curve, which can be understood that the $h_2$ mass is cancelled out in the final expression in Eq.~(\ref{DDXS}) in this parameter region. However, even though it is remarkable that the LUX upper limit of $\kappa$ reaches the order of $10^{-10}$ for large VDM masses, it is not able to give a meaningful constraints on the freeze-in region of our model. Thus, in the following, we will not consider the direct detection constraints any more.

\section{Indirect Detection Constraints on Vector Dark Matters}\label{Sec_ID}
\begin{figure}[ht]
\includegraphics[scale = 0.8]{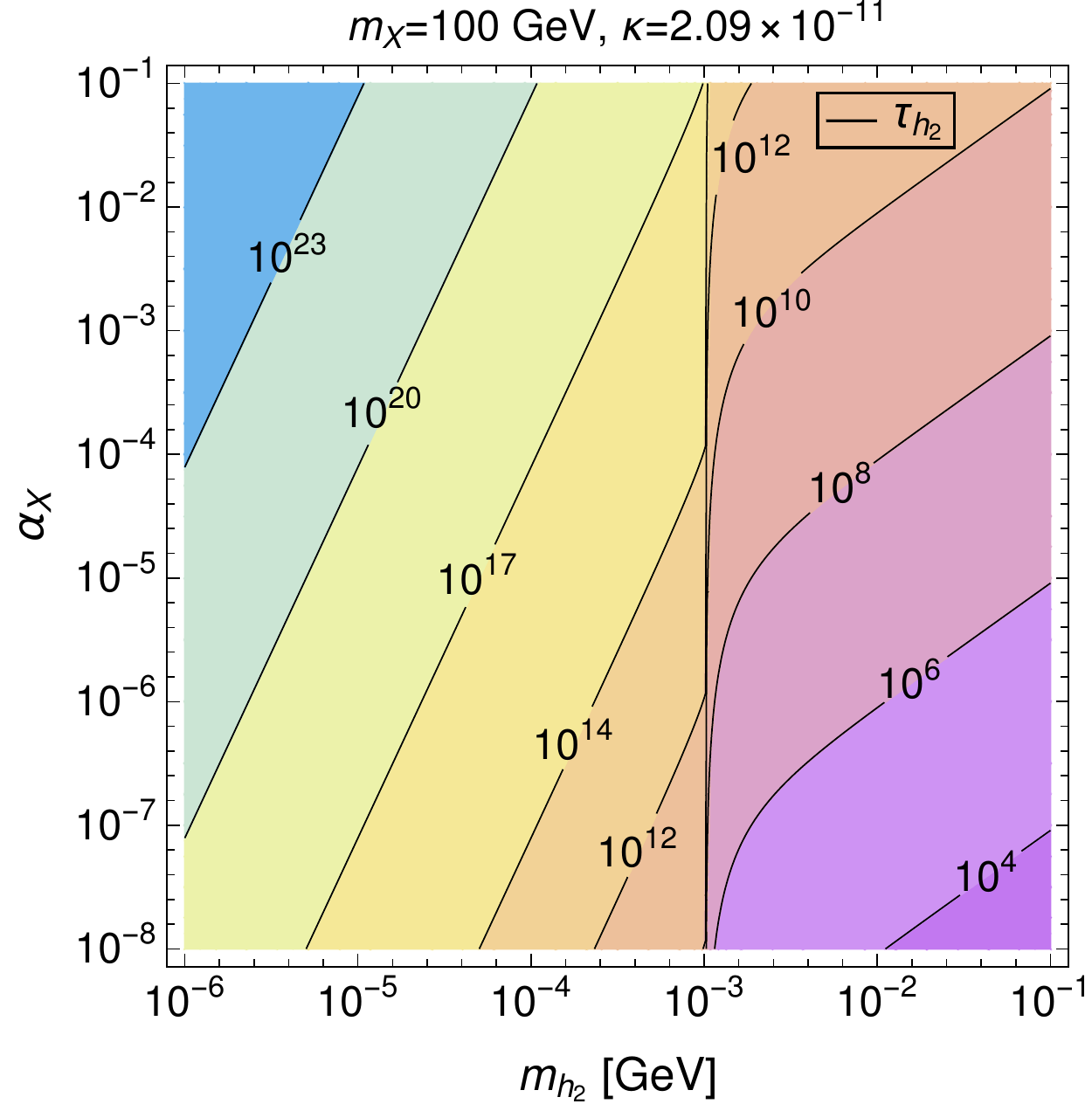}
\caption{Contour plot for the $h_2$ lifetime in the $m_{h_2}$-$\alpha_X$ plane for VDM with $m_X = 100$~GeV. The Higgs portal coupling $\kappa = 2.09\times 10^{-11}$ is determined by the VDM relic density through the freeze-in mechanism. The numbers on the line represent the $h_2$ lifetime in units of seconds. }\label{Fig_aX}
\end{figure}

Phenomenology of indirect detection of VDM crucially depends on properties of the mediator $h_2$, such as its mass $m_{h_2}$, lifetime $\tau_{h_2}$ and dominant decay channels. Since we are interested in the light $h_2$ which could give rise to the large enhancement of VDM self-interactions, we will limit ourself to $m_{h_2} \lesssim 100$~MeV. Thus, the parameter space is naturally divided into two regions: (i) $m_{h_2} \geq 2 m_e$ and (ii) $m_{h_2} < 2 m_e$, where $m_e$ is the electron mass. In the former region, the dominant $h_2$ decay channel is $e^+ e^-$ pairs, while only the diphoton mode is kinematically available in the latter case. Consequently, the light mediator lifetime $\tau_{h_2}$ is different in these two regions. Specifically, $10^4~{\rm s} \lesssim \tau_{h_2} \lesssim 10^{12}~{\rm s}$ in region (i) while $\tau_{h_2} \gtrsim 10^{12}~{\rm s}$ in region (ii), which is illustrated in Fig.~\ref{Fig_aX} for a typical VDM mass $m_X= 100$~GeV and a Higgs portal coupling $\kappa=2.09 \times 10^{-11}$ consistently with the DM relic density (see Fig.~\ref{RelicDensity}). Analyzing constraints from DM indirect searches, we will consider these two regions separately.

\noindent\textbf{Region (i)}: $2 m_e \lesssim m_{h_2} \lesssim 100$~MeV

Since $h_2$ dominantly decays into $e^+ e^-$ pairs resulting in its lifetime of $10^4~{\rm s} \lesssim \tau_{h_2} \lesssim 10^{12}$~s, the relevant indirect detection constraints involve the following experiments:

\begin{itemize}
\item \textbf{Big Bang Nucleosynthesis (BBN)}: Due to its long lifetime $\tau_{h_2} \gtrsim 10^4$~s, the dark Higgs boson $h_2$  was still present in the early Universe at the epoch of BBN. Thus from the viewpoint of BBN, $h_2$ plays the role of an extra, decaying component of DM. Such a late decay of $h_2$ would produce $e^+ e^-$ pairs with sufficient energy that would spoil the predictions of abundances of various elements~\cite{Jedamzik:2009uy,Berger:2016vxi,Kawasaki:2004qu,Kawasaki:2017bqm}. We adopt the most recent results from Ref.~\cite{Berger:2016vxi} where the authors also studied the BBN effects triggered by decays of dark Higgs bosons produced by the freeze-in mechanism. It is seen from Fig.~8 in Ref.~\cite{Berger:2016vxi} that the most stringent constraint to the model is set by $s_{\theta}<5\times 10^{-12}$ for $1~{\rm MeV} < m_{h_2}< 100~{\rm MeV}$ where $\theta$ is the mixing angle defined in Eq.~(\ref{ScalarTrans}). On the other hand, when $m_{h_2} < 1$~MeV, there is no constraints to the decaying $h_2$ at all. Note that the result in Ref.~\cite{Berger:2016vxi} was obtained in the limit of $\kappa\to 0$ and $v_S \to \infty$ while keeping $\theta$ fixed, so the $2 \to 2$ processes involving top quarks predominate the $h_2$ production via freeze-in. However, in our scenario, the Higgs portal coupling does not approach zero. The most important contribution to $h_2$ density arises from the SM-like Higgs decay $h_1 \to h_2 h_2$, which is more efficient than the top quark annihilations and top-gluon inelastic scatterings. Therefore, we expect that $h_2$ is more abundant in the our model, which leads to even stronger constraints. In other words, the application of dark Higgs results in Ref.~\cite{Berger:2016vxi} here leads to the conservative constraints.

\item \textbf{Cosmic Microwave Background (CMB)}: The CMB was formed when the photon last scattering occurred at $\sim 10^{12}$~s after the Big Bang.
For most of the parameter space the duration of CMB formation ($\sim 10^{12}$~s) was much longer than the $h_2$ lifetime. Thus annihilation of the VDMs into a pair of $h_2$~\cite{Ade:2015xua}, which further decayed  into energetic electrons and positrons, distorted the CMB spectrum in Cosmic Dark Ages~\cite{Padmanabhan:2005es,Slatyer:2015jla,Slatyer:2015kla}. Moreover, such a CMB constraint was further strengthened by the Yukawa potential in Eq.~(\ref{potential}) via the Sommerfeld enhancement~\cite{Sommerfeld,ArkaniHamed:2008qn}. For the $s$-wave VDM annihilations, this can lead to a nonperturbative correction to the tree-level cross section, $\sigma v = S \times (\sigma v)_0$, in which $(\sigma v)_0$ denotes the tree-level perturbative cross section for $X X \to h_2 h_2$ and $S$ is the $s$-wave Sommerfeld enhancement factor given by~\cite{Tulin:2013teo,Cassel:2009wt,Iengo:2009xf,Slatyer:2009vg}
\begin{eqnarray}\label{SomEnh}
S = \frac{\pi}{a} \frac{\sinh(2\pi a c)}{\cosh(2\pi a c)-\cos(2\pi\sqrt{c-(ac)^2})}\,,
\end{eqnarray}
with $a \equiv v/(2\alpha_X)$ and $c\equiv 6\alpha_X m_X/(\pi^2 m_{h_2})$. Since the velocity of the VDM was very small during the photon last scattering, we can use the value of $S$ saturated in the vanishing velocity limit. Due to the large mass hierarchy between the VDM $X$ and the mediator $h_2$, the CMB upper limit in Fig.~8 of Ref.~\cite{Elor:2015bho} for the one-step cascade with the $e^+ e^-$ final state can be applied for the VDM annihilation cross-section.

\item \textbf{AMS-02}: The local annihilations of VDMs into $h_2$ pairs decaying to $e^+ e^-$ in the final state can lead to an excess of positron flux in cosmic rays~\cite{Bergstrom:2013jra,Hooper:2012gq,Ibarra:2013zia}. Therefore, the absence of such an excess would give rise to a strong upper bound on the VDM annihilation cross section. Currently, the most precise measurements of the positron flux~\cite{AMS1} and positron fraction~\cite{AMS2} come from the AMS-02 Collaboration. By taking into the account the Sommerfeld enhancement factor in Eq.~(\ref{SomEnh}) with typical VDM velocity $v_X \sim 10^{-3}$ in our Galaxy, we can take the AMS-02 positron flux constraints from Ref.~\cite{Elor:2015bho} for one-step cascading VDM annihilations. Note that the AMS-02 results are reliable only down to the DM mass $\sim 10$~GeV, since the positron flux spectrum lower than 10~GeV would be affected significantly by the solar modulation so that the constraints in this range would be uncertain.

\item \textbf{Dwarf Limits from Fermi}: The VDM annihilations in the dwarf spheroidal galaxies provide bright $\gamma$-ray sources in the Milky Way, and are thus expected to be probed and constrained by the Fermi Gamma-Ray Space Telescope~\cite{Ackermann:2015zua}. In the present model with $m_{h_2}> 2 m_e$, most $\gamma$-rays are generated by the final-state radiation from the mediator decay $h_2 \to e^+ e^- \gamma$, which follows the VDM annihilation $X X \to h_2 h_2$. However, due to the suppression factor from radiative corrections compared with the dominant decay channel $h_2 \to e^+ e^-$, the constraints from Fermi shown in Ref.~\cite{Elor:2015bho} are much weaker than the corresponding ones from CMB and AMS-02. Therefore, we do not show dwarf limits from Fermi in our following numerical results.

\end{itemize}

\noindent\textbf{Region (ii)}: $m_{h_2} < 2 m_e$

We now turn to the indirect search constraints for the VDM with the mediator mass $m_{h_2} < 2 m_e$, in which $h_2$ decays dominantly in the diphoton channel, and the lifetime is typically longer than $10^{12}$~s. As mentioned before, for such a light $h_2$, the BBN constraints can be evaded as shown in Ref.~\cite{Berger:2016vxi}.

\begin{itemize}
\item \textbf{Dwarf Limits from Fermi}: Since $h_2 \to \gamma\gamma$ is the dominant $h_2$ decay we expect that there should be strong constraints from measurements of $\gamma$-rays by Fermi Gamma-Ray Space Telescope~\cite{Ackermann:2015zua}. However, note that the signal region for each dwarf is defined as the one within an angular radius of $0.5^{\circ}$. For the 15 dwarfs used in the Fermi-LAT analysis, their distances from the Earth range from 32 kpc to 233 kpc. Thus, due to the fact that $h_2$ propagates at the speed of light without any scatterings in the range of a dwarf, the $h_2$ will spend, at most, the time of ${\cal O}(10^{11}~{\rm s})$ traveling inside the signal region from the center of the dwarf. In other words, it is too short in time for $h_2$ to decay inside the signal region. As a result, the Fermi-LAT constraints in Ref.~\cite{Ackermann:2015zua} cannot be adopted directly in our case.

\item \textbf{CMB}: For $\tau_{h_2} > 10^{12}$~s, $h_2$ would have a large abundance at the time of recombination. Also, the high-energy photons from $h_2$ decays would ionize and heat neutral hydrogen after recombination, and hence distort the CMB anisotropy spectrum. Consequently, recent measurements of the CMB by Planck~\cite{Ade:2015xua} can provide strong constraints on $h_2$ properties~\cite{Adams:1998nr,Chen:2003gz}. We adopt the recent lower bound on the decaying DM lifetime $\tau^0$ for the diphoton final state shown in Fig.~7 of  Ref.~\cite{Slatyer:2016qyl} to obtain the following constraint for the VDM model:
\begin{eqnarray}\label{CMBd}
\tau_{h_2} \geq \tau^0 \times \frac{\Omega_{h_2} h^2}{0.12}\,,
\end{eqnarray}
where $\Omega_{h_2}h^2$ is the current $h_2$ relic density generated via the freeze-in mechanism if $h_2$ were present today without decays. In fact, $h_2$ might decay well before. The constraint is actually for $h_2$ abundance at the epoch of recombination, not today. We only use the present DM relic density as a reference to quantify the $h_2$ density fraction at the recombination period. Moreover, the expression on the right-hand side in Eq.~(\ref{CMBd}) is just an approximation and the true formula should be $\Omega_{h_2}h^2/(\Omega_{h_2}h^2+\Omega_X h^2)$. However, the $h_2$ density is always smaller than that of VDM due to the assumed mass hierarchy, so that $\Omega_{h_2}h^2$ in the denominator can be neglected. Note that the exclusion limit in Ref.~\cite{Slatyer:2016qyl} extends to the DM mass of 10~keV, so we ignore the CMB constraints below this VDM mass in our numerical calculations.

\item \textbf{Diffuse $\gamma$/$X$-Ray Bounds}: When the lifetime of $h_2$ is larger than the present age of the Universe $\tau_U = 4.3\times 10^{17}$~s, the $h_2$ particle contributes to the present DM relic density even though it is not absolutely stable. The only decay channel $h_2 \to \gamma\gamma$ could be constrained by the accurate measurement of the diffuse $\gamma$/$X$-ray background. Following Ref.~\cite{Bernal:2015ova,Essig:2013goa,Boddy:2015efa,Riemer-Sorensen:2015kqa}, we adopt the conservative lower limit on the $h_2$ lifetime as
\begin{eqnarray}
\tau_{h_2} \gtrsim 10^{28}~{\rm sec} \times \frac{\Omega_{h_2}h^2}{0.12}\,,
\end{eqnarray}
where $\Omega_{h_2}$ is the relic abundance of $h_2$ generated via the freeze-in mechanism.

\end{itemize}


\section{Numerical Results}\label{Sec_Res}
\begin{figure}[ht]
\includegraphics[scale = 0.39,angle=0]{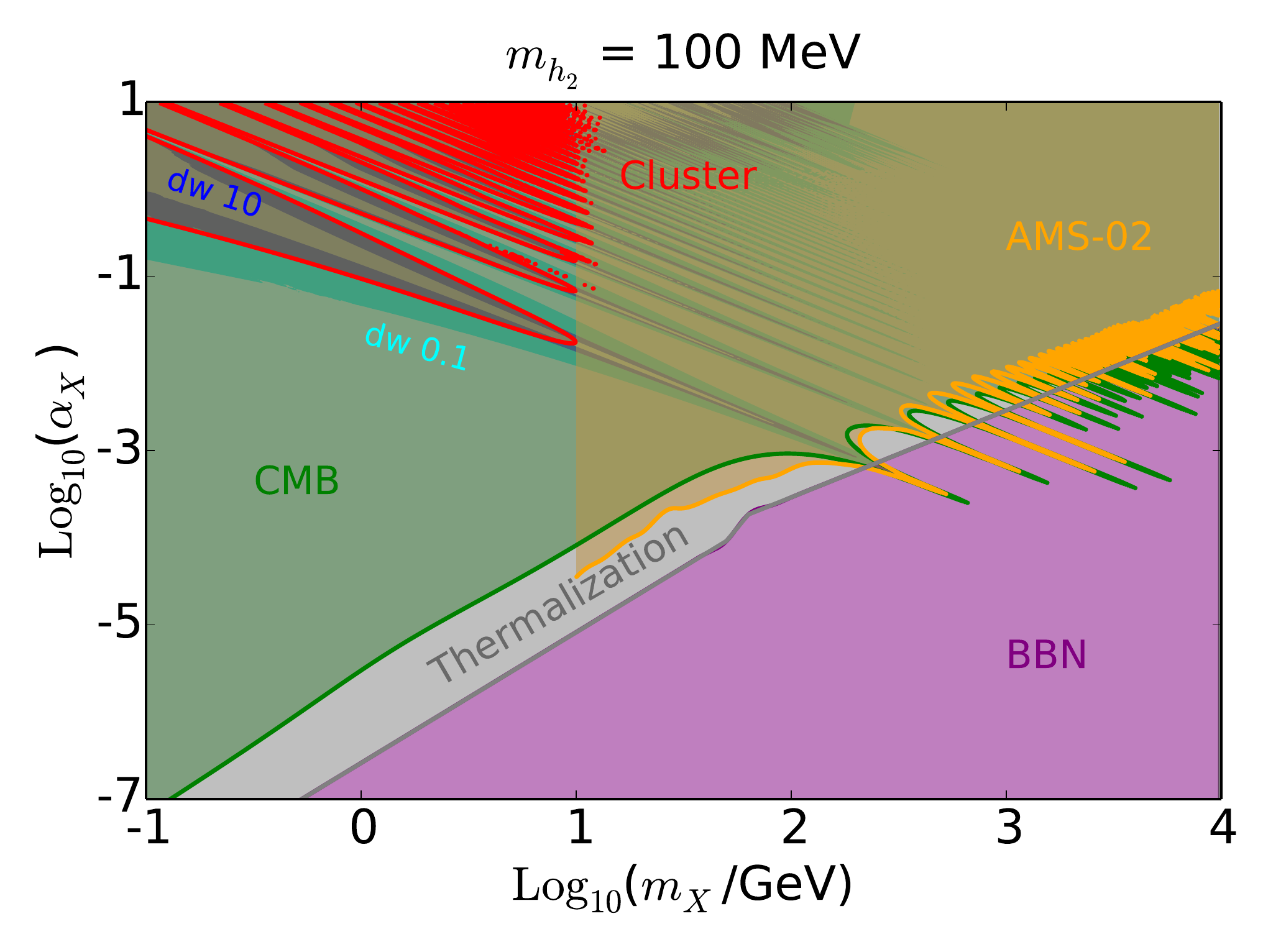}
\includegraphics[scale = 0.39,angle=0]{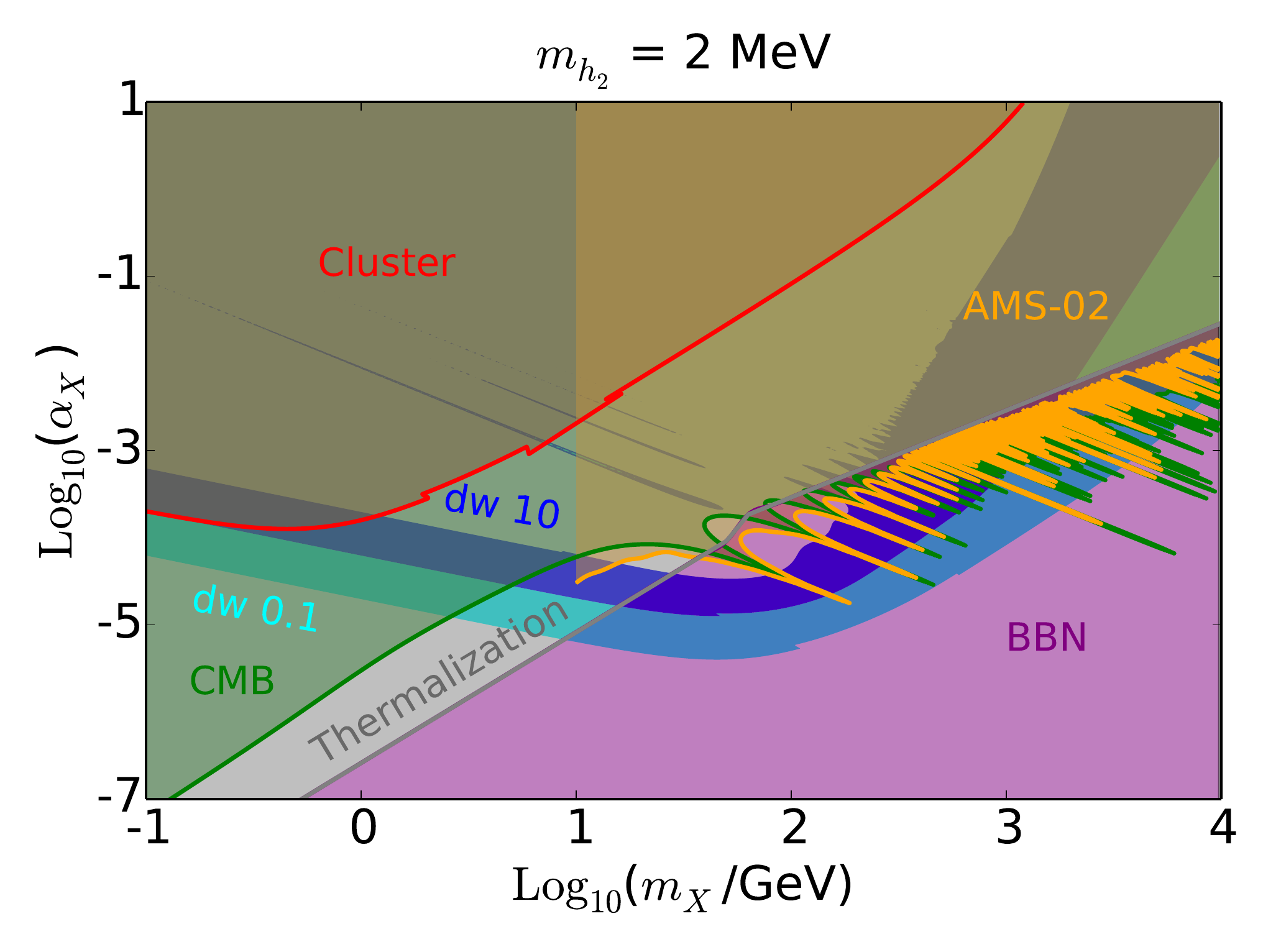}
\caption{Region (i) in the parameter space with $m_{h_2} = 100$~MeV (left panel) and $m_{h_2} = 2$~MeV (right panel) for which the VDM self scattering might be sufficiently strong to solve the small scale structure anomalies. The numbers next to "dw" represent $\sigma_T/m_X$ in units of ${\rm cm}^2/g$ at the dwarf scale. The blue (cyan) region shows the space with the VDM self-interaction cross section of $1~{\rm cm^2/g} < \sigma_T/m_X < 10~{\rm cm^2/g}$ ($0.1~{\rm cm^2/g} < \sigma_T/m_X < 1~{\rm cm^2/g}$) at the dwarf scale, while the red region shows bounds on VDM self scatterings at the cluster scale. The purple, green, and orange colors show regions excluded by the DM indirect searches from BBN, CMB, and AMS-02, respectively. The region below the curve named ``Thermalization'' shows the parameter space in which the VDM is generated via freeze-in, and the gray region above the curve corresponds to the one with dark sector thermalized in itself. In both plots the Higgs portal coupling $\kappa$ is fixed by the VDM relic density as a function of $m_X$  in the freeze-in region. }\label{Fig_DMISi}
\end{figure}

\begin{figure}[ht]
\includegraphics[scale = 0.39,angle=0]{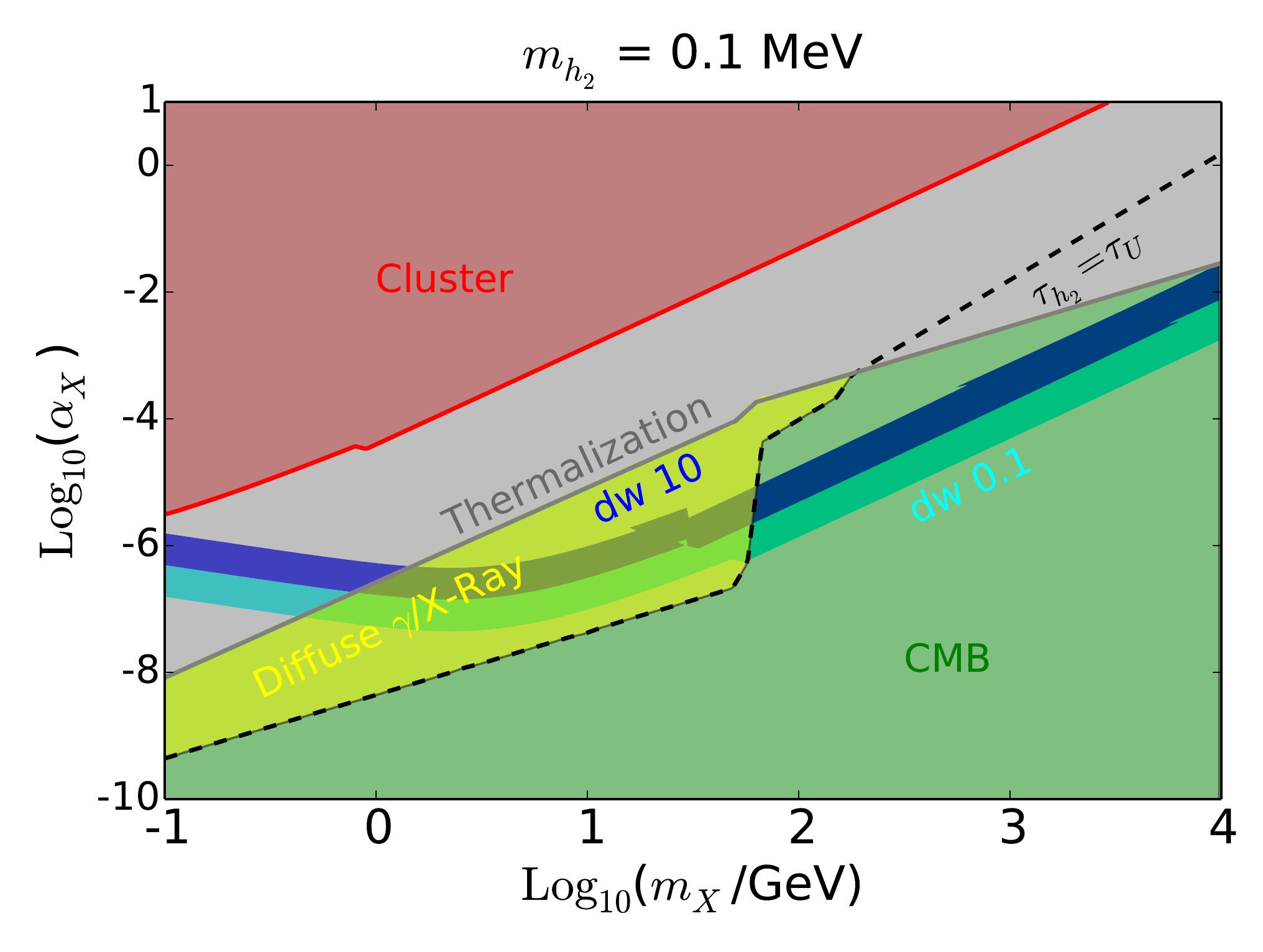}
\includegraphics[scale = 0.39,angle=0]{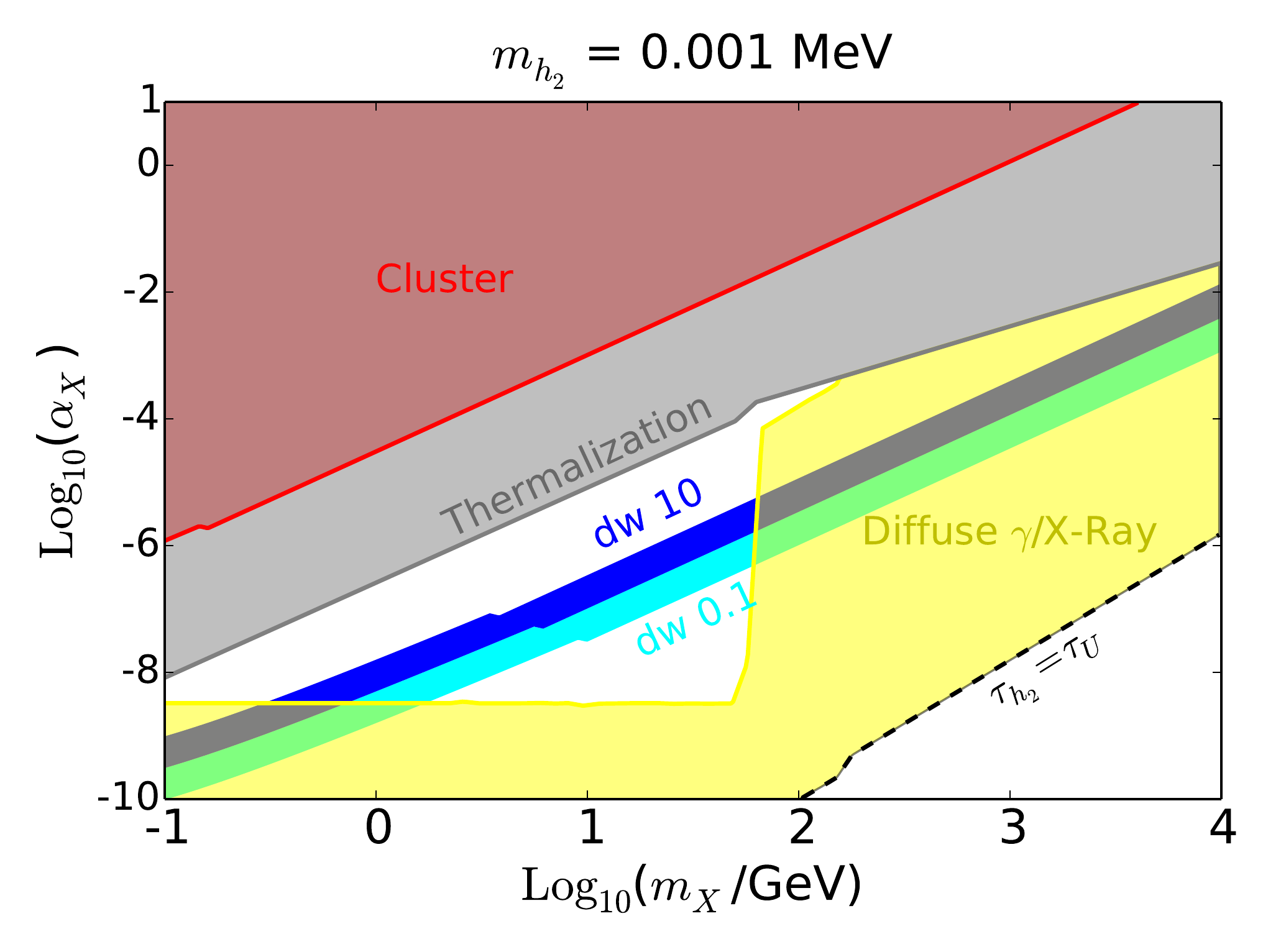}
\caption{Same as in Fig.~\ref{Fig_DMISi} but for Region (ii) with $m_{h_2} = 10^{-1}$ MeV (left panel) and $m_{h_2} = 10^{-3}$ MeV (right panel). When the lifetime of $h_2$ is longer than the current age of the Universe $\tau_U = 4.3 \times 10^{17}$~s (denoted by the dashed black curve), we also include the constraints from the diffuse $\gamma$/$X$-ray observations. }\label{Fig_DMIS2}
\end{figure}

Having discussed all of the VDM signals and constraints, we can put everything together to see if we can find a region in the  parameter space where large DM self-scatterings for the scale of dwarf galaxies can be compatible with the VDM relic density and all of indirect search constraints. Note that there are four free parameters in our original VDM model, so that if the Higgs portal coupling $\kappa$ is fixed as shown in Fig.~\ref{RelicDensity} by the requirement that the VDM relic density constitute all of the DM in the Universe, we can plot the parameter space in the $m_{X}$-$\alpha_X$ plane with fixed values of $m_{h_2}$. The final results for some typical values of $m_{h_2}$ in the Regions (i) and (ii) are presented in Fig.~\ref{Fig_DMISi} and Fig.~\ref{Fig_DMIS2}, respectively.

Fig.~\ref{Fig_DMISi} shows constraints on the parameter space for two extremal values of $m_{h_2}$ in the Region (i). {Note that only the region below the curve named ``Thermalization'' represents the freeze-in region, so that above the curve the parameter space corresponds to the case with dark sector thermalized by itself.
It is seen that the DM indirect search constraints from BBN, CMB, and AMS-02 are so strong that the whole freeze-in region is excluded, no matter how precisely we tune the model parameters.} If we focus on the region that solves the small-scale structure problems, the left panel shows that {the only parameters that exactly sit at the resonances can enter the freeze-in region
for a relatively heavy ${h_2}$ ($m_{h_2} = 100$~GeV).} In this case, the most stringent constraints come from the CMB and AMS-02, which can be understood that the same resonances that give rise to the appropriate VDM self-interactions would also induce large Sommerfeld enhancements in the VDM annihilations. However, as the $h_2$ mass decreases, more and more signal regions at the dwarf scale are shifted to the classical and Born regions. In the extremal case with $m_{h_2} = 2$~MeV which is chosen to avoid the $e^+ e^-$ threshold effect around $1$~MeV, the velocity dependence of the VDM self-scatterings becomes manifest, as the constrained region from clusters separates from the signal region at the dwarf scale. However, the whole signal region below the thermalization curve is excluded by the combination of the CMB, AMS-02 and BBN constraints.

The situation changes a lot for the Region (ii) as shown in Fig.~\ref{Fig_DMIS2}, since the indirect detection constraints are all imposed on the decay process $h_2 \to \gamma\gamma$, rather than VDM annihilations. Here we only consider the freeze-in region below the thermalization curve. In both panels, the signal regions for the dwarf galaxy scale are all in the Born and classical regions, in which the part with a small hidden gauge coupling $\alpha_X$ and a light VDM corresponds to the Born region while the band with large values of $\alpha_X$ and $m_X$ to the classical region. The discontinuities in both plots represent the mismatch of the analytical formulae around the boundary of these two regions. By detailed calculations, it is found that all of the signal regions for $m_{h_2} \gtrsim 10^{-2}~$MeV are constrained tightly by observations of CMB and diffuse $\gamma$/$X$-rays, as illustrated by the left panel of Fig.~\ref{Fig_DMIS2}. Only when the $h_2$ mass is reduced to ${\cal O}$(keV) a small parameter window opens, in which the VDM mass is around ${\cal O}({\rm GeV})$ and $\alpha_X$ is in the range $10^{-9}\sim 10^{-6}$, as is seen clearly in the right panel of Fig.~\ref{Fig_DMIS2}.

\section{Conclusion}\label{Sec_Conc}
We have investigated a simple VDM model in which the dark sector consists of only two particles, the VDM $X$ and the dark Higgs $h_2$, and it couples to the SM sector weakly through the Higgs portal. We are particularly interested in the region of the parameter space with the VDM mass of ${\cal O}({\rm GeV} \sim {\rm TeV})$ and the $h_2$ mass of $m_{h_2} < 100$~MeV, where the dark Higgs $h_2$ plays a role of a light mediator so that non-perturbative effects can generate VDM self-interactions with the appropriate magnitude to solve the small-scale structure problems at the dwarf galaxy scale. Due to the velocity dependence, such VDM self-scatterings can avoid the constraints at the galaxy cluster scale. In our work, we consider the scenario in which the VDM relic density is produced by the freeze-in mechanism. Especially, we pay attention to consequences of the EW phase transition, since they affect the dominant DM production channels quite dramatically. It turns out that the Higgs portal coupling $\kappa$ is always predicted to be of ${\cal O}(10^{-11})$. With such a tiny portal coupling, the DM direct detection limits cannot constrain the model much, as it is illustrated by the latest complete LUX data.

However, the indirect detections can place strong constraints on the parameter space of interest. Specifically, when $2 m_e< m_{h_2} < 100$~MeV, $h_2$ decays dominantly via $h_2 \to e^+ e^-$ so that the limits from BBN, CMB, and AMS-02 exclude all of the parameter space. In particular, the CMB and AMS-02 bounds are strengthened since the main VDM annihilation $XX \to h_2 h_2$ is enhanced by the non-perturbative Sommerfeld effects. On the other hand, if $m_{h_2}<2m_e$, the relevant indirect search constraints come from observations of CMB and diffuse $\gamma$/X-rays. As a result, only when the mass of $h_2$ is equal to or smaller than keV scale one can find a parameter window in which the model can lead to the right DM relic abundance and appropriate DM self-interactions, while it does not conflict with other indirect detection observations.

\appendix
\section{Relevant Cross Sections and Decay Rates for the Higgs Portal}\label{appA}
In this appendix, we collect the formulae for the relevant SM particle annihilation cross sections as well as SM-like Higgs $h_1$ decay rate, which are involved in the calculation of dark matter relic density in the Universe via the freeze-in mechanism. Since the SM EW phase transition has a substantial impact on the VDM production, we consider the annihilation and decay channels in broken and symmetric phases, respectively. \\

\noindent {\bf EW Symmetry-Broken Phase:}

\underline{Quark Annihilation to VDMs}:
\begin{eqnarray}
\sigma (q\bar{q}\to X X) =  \frac{\kappa^2 m_q^2}{192\pi}\sqrt{\frac{(s-4 m_q^2)(s-4 m_X^2)}{s^2}} \frac{s^2-4 m_X^2 s + 12 m_X^4}{(s-m_{h_2}^2)^2 [(s-m_{h_1}^2)^2+\Gamma_{h_1}^2 m_{h_1}^2]}\,,
\end{eqnarray}
where we use the SM Higgs boson width $\Gamma_h=4.15$~MeV~\cite{Dittmaier:2011ti, Heinemeyer:2013tqa} to regulate the SM-like Higgs mass pole singularity.

\underline{Lepton Annihilation to VDMs}:
\begin{eqnarray}
\sigma (l\bar{l}\to X X) =  \frac{\kappa^2 m_l^2}{64\pi}\sqrt{\frac{(s-4 m_l^2)(s-4 m_X^2)}{s^2}} \frac{s^2-4 m_X^2 s + 12 m_X^4}{(s-m_{h_2}^2)^2 [(s-m_{h_1}^2)^2+\Gamma_{h_1}^2 m_{h_1}^2]}\,,
\end{eqnarray}

\underline{$W$ Bosons Annihilation to VDMs}:
\begin{eqnarray}
\sigma (W^+ W^-\to X X) &=&  \frac{\kappa^2}{288\pi}\sqrt{\frac{(s-4 m_W^2)(s-4 m_X^2)}{s^2}} \frac{s^2-4 m_X^2 s + 12 m_X^4}{(s-m_{h_2}^2)^2 [(s-m_{h_1}^2)^2+\Gamma_{h_1}^2 m_{h_1}^2]}\nonumber\\
&& \times \frac{s^2-4 m_W^2 s + 12 m_W^4}{s-4 m_W^2}\,,
\end{eqnarray}

\underline{$Z$ Bosons Annihilation to VDMs}:
\begin{eqnarray}
\sigma (Z Z\to X X) &=&  \frac{\kappa^2}{288\pi}\sqrt{\frac{(s-4 m_Z^2)(s-4 m_X^2)}{s^2}} \frac{s^2-4 m_X^2 s + 12 m_X^4}{(s-m_{h_2}^2)^2 [(s-m_{h_1}^2)^2+\Gamma_{h_1}^2 m_{h_1}^2]}\nonumber\\
&& \times \frac{s^2-4 m_Z^2 s + 12 m_Z^4}{s-4 m_Z^2}\,,
\end{eqnarray}

\underline{SM-Like Higgs Bosons $h_1$ Annihilaiton to VDMs}:
\begin{eqnarray}
\sigma (h_1 h_1\to X X) &\simeq &  \frac{\kappa^2}{32\pi}\sqrt{\frac{(s-4 m_{h_1}^2)(s-4 m_X^2)}{s^2}} \frac{s^2-4 m_X^2 s + 12 m_X^4}{(s-m_{h_2}^2)^2 [(s-m_{h_1}^2)^2+\Gamma_{h_1}^2 m_{h_1}^2]}\nonumber\\
&& \times \frac{(s+2m_{h_1}^2)^2}{s-4 m_{h_1}^2}\,,
\end{eqnarray}
where we have only kept the leading-order terms in the double expansion of $\kappa$ and $m^2_{h_2}/m^2_X$.

\underline{SM-Like Higgs Boson $h_1$ Decay to VDMs}:

When the VDM mass is smaller than a half of the the SM-like Higgs mass, the VDM can also be produced by the decay of $h_1$, with the decay rate as follows:
\begin{eqnarray}
\Gamma(h_1 \to X X) &=& \frac{\alpha_X s^2_\theta}{2} \sqrt{m_{h_1}^2-4m_X^2} \frac{m_X^2}{m_{h_1}^2} \left[2+\frac{(m_{h_1}^2-2m_X^2)^2}{4m_X^4}\right]\nonumber\\
&\simeq &  \frac{2\kappa^2 s_W^2 c_W^2}{\alpha} \frac{m_Z^2 m_X^4}{(m_{h_1}^2-m_{h_2}^2)^2 m_{h_1}^2}  \sqrt{m_{h_1}^2-4m_X^2} \left[2+\frac{(m_{h_1}^2-2m_X^2)^2}{4m_X^4}\right]\,,
\end{eqnarray}
where we have used the definition of $\kappa$ in Eq.~(\ref{def_k}) and the approximation that $c_\theta \approx 1$.

\underline{SM-like Higgs Boson $h_1$ Decay into $h_2$'s}:
\begin{eqnarray}
\Gamma(h_1 \to h_2 h_2) \approx \frac{\kappa^2 v_H^2}{32\pi m_{h_1}}\,,
\end{eqnarray}
where we only keep the leading order in the expansion of $\kappa$ and $m_{h_2}/m_{h_1}$ because $m_{h_2} \ll m_{h_1}$.

\noindent {\bf EW Symmetric Phase: }

\underline{SM Higgs Doublet $H$ Annihilation to VDMs}
\begin{eqnarray}
\sigma(X X \to H H^\dagger) = \frac{\kappa^2}{72\pi} \sqrt{\frac{(s-4 m_{H}^2)(s-4m_X^2)}{s^2}} \frac{s^2-4 m_X^2 s + 12 m_X^4}{(s-m_s)^2(s-4m_X^2)}\,,
\end{eqnarray}
where $m_H$ and $m_s$ are the masses of the SM-Higgs doublet and the dark Higgs $\phi_S$ defined in Eq.~(\ref{ScalarBEW}).

Note that in our derivation of the Boltzmann equation in Eq.~(\ref{BEq}) after and before the EW phase transition, we have used the so-called reaction density $\gamma_i$ for various channels. For the annihilation channels, the reaction density is defined as~\cite{Edsjo:1997bg}
\begin{eqnarray}
\gamma (a \, b \to 1\, 2) &\equiv & \int  d\bar{p}_a d\bar{p}_b   d\bar{p}_1 d\bar{p}_2 f_a^{\rm eq} f_b^{\rm eq} (2\pi)^4 \delta^4(p_a +p_b-p_1-p_2) |{\cal M}(a\, b\to 1\,2)|^2\nonumber\\
&=& \frac{T}{32\pi^4} g_a g_b \int^\infty_{s_{\rm min}}ds \frac{[(s-m_a^2-m_b^2)^2-4m_a^2 m_b^2]}{\sqrt{s}} \sigma(a\,b \to 1\,2) K_1\left(\frac{\sqrt{s}}{T}\right)\,,\nonumber\\
\end{eqnarray}
where $a,b$ ($1,2$) represent the incoming (outgoing) particles with $g_{a,b}$ as their respective degrees of freedom, and $f_i^{\rm eq} \approx e^{-E_i/T}$ is the Maxwell-Boltzmann distribution. Here $d\bar{p} \equiv d^3p/(2\pi)^3 (2E)$, $|{\cal M}|^{2}$ are the amplitude squared summed over quantum numbers of the initial and final states without averaging, and $s_{\rm min} = {\rm max} [(m_a+m_b)^2,(m_1+m_2)^2]$.

On the other hand, the reaction density for the decay channel $a \to 1 2$ has the following definition~\cite{Hall:2009bx,Chu:2011be}:
\begin{eqnarray}
\gamma^D({a \to 1\, 2})  &\equiv & \int d \bar{p}_{a} d \bar{p}_{1} \bar{p}_2 (2\pi)^4 \delta^4(p_{a}-p_1-p_2)  f^{\rm eq}_{a} |{\cal M}(a \to 1\,2)|^2\nonumber\\
& = & \frac{g_a}{2\pi^2} m_a^2 \Gamma(a\to 1\,2) T K_1\left(\frac{m_a}{T}\right)\,,
\end{eqnarray}
where $\Gamma_{a\to 1\,2}$ is the zero-temperature decay rate.

\section*{Acknowledgments}
We would like to thank Kai Schmidt-Hoberg, Laura Covi, and Chao-Qiang Geng for useful discussions. This work is supported by the National Science Centre (Poland) research project, decision DEC-2014/15/B/ST2/00108.



\begin{thebibliography}{0}
\bibitem{PDG}
  C.~Patrignani {\it et al.} [Particle Data Group],
  Chin.\ Phys.\ C {\bf 40}, no. 10, 100001 (2016).
  doi:10.1088/1674-1137/40/10/100001

\bibitem{Bergstrom:2012fi}
  L.~Bergstrom,
  Annalen Phys.\  {\bf 524}, 479 (2012)
  doi:10.1002/andp.201200116
  [arXiv:1205.4882 [astro-ph.HE]].

\bibitem{Moore:1994yx}
  B.~Moore,
  Nature {\bf 370}, 629 (1994).
  doi:10.1038/370629a0

\bibitem{Flores:1994gz}
  R.~A.~Flores and J.~R.~Primack,
  Astrophys.\ J.\  {\bf 427}, L1 (1994)
  doi:10.1086/187350
  [astro-ph/9402004].

\bibitem{Oh:2010ea}
  S.~H.~Oh, W.~J.~G.~de Blok, E.~Brinks, F.~Walter and R.~C.~Kennicutt, Jr,
  Astron.\ J.\  {\bf 141}, 193 (2011)
  doi:10.1088/0004-6256/141/6/193
  [arXiv:1011.0899 [astro-ph.CO]].

\bibitem{Walker:2011zu}
  M.~G.~Walker and J.~Penarrubia,
  Astrophys.\ J.\  {\bf 742}, 20 (2011)
  doi:10.1088/0004-637X/742/1/20
  [arXiv:1108.2404 [astro-ph.CO]].


\bibitem{BoylanKolchin:2011de}
  M.~Boylan-Kolchin, J.~S.~Bullock and M.~Kaplinghat,
  Mon.\ Not.\ Roy.\ Astron.\ Soc.\  {\bf 415}, L40 (2011)
  doi:10.1111/j.1745-3933.2011.01074.x
  [arXiv:1103.0007 [astro-ph.CO]].

\bibitem{BoylanKolchin:2011dk}
  M.~Boylan-Kolchin, J.~S.~Bullock and M.~Kaplinghat,
  Mon.\ Not.\ Roy.\ Astron.\ Soc.\  {\bf 422}, 1203 (2012)
  doi:10.1111/j.1365-2966.2012.20695.x
  [arXiv:1111.2048 [astro-ph.CO]].

\bibitem{Garrison-Kimmel:2014vqa}
  S.~Garrison-Kimmel, M.~Boylan-Kolchin, J.~S.~Bullock and E.~N.~Kirby,
  Mon.\ Not.\ Roy.\ Astron.\ Soc.\  {\bf 444}, no. 1, 222 (2014)
  doi:10.1093/mnras/stu1477
  [arXiv:1404.5313 [astro-ph.GA]].


\bibitem{deLaix:1995vi}
  A.~A.~de Laix, R.~J.~Scherrer and R.~K.~Schaefer,
  Astrophys.\ J.\  {\bf 452}, 495 (1995)
  doi:10.1086/176322
  [astro-ph/9502087].

\bibitem{Spergel:1999mh}
  D.~N.~Spergel and P.~J.~Steinhardt,
  Phys.\ Rev.\ Lett.\  {\bf 84}, 3760 (2000)
  doi:10.1103/PhysRevLett.84.3760
  [astro-ph/9909386].

\bibitem{Vogelsberger:2012ku}
  M.~Vogelsberger, J.~Zavala and A.~Loeb,
  Mon.\ Not.\ Roy.\ Astron.\ Soc.\  {\bf 423}, 3740 (2012)
  doi:10.1111/j.1365-2966.2012.21182.x
  [arXiv:1201.5892 [astro-ph.CO]].

\bibitem{Zavala:2012us}
  J.~Zavala, M.~Vogelsberger and M.~G.~Walker,
  Mon.\ Not.\ Roy.\ Astron.\ Soc.\  {\bf 431}, L20 (2013)
  doi:10.1093/mnrasl/sls053
  [arXiv:1211.6426 [astro-ph.CO]].

\bibitem{Rocha:2012jg}
  M.~Rocha, A.~H.~G.~Peter, J.~S.~Bullock, M.~Kaplinghat, S.~Garrison-Kimmel, J.~Onorbe and L.~A.~Moustakas,
  Mon.\ Not.\ Roy.\ Astron.\ Soc.\  {\bf 430}, 81 (2013)
  doi:10.1093/mnras/sts514
  [arXiv:1208.3025 [astro-ph.CO]].

\bibitem{Peter:2012jh}
  A.~H.~G.~Peter, M.~Rocha, J.~S.~Bullock and M.~Kaplinghat,
  Mon.\ Not.\ Roy.\ Astron.\ Soc.\  {\bf 430}, 105 (2013)
  doi:10.1093/mnras/sts535
  [arXiv:1208.3026 [astro-ph.CO]].

\bibitem{Kaplinghat:2015aga}
  M.~Kaplinghat, S.~Tulin and H.~B.~Yu,
  Phys.\ Rev.\ Lett.\  {\bf 116}, no. 4, 041302 (2016)
  doi:10.1103/PhysRevLett.116.041302
  [arXiv:1508.03339 [astro-ph.CO]].

\bibitem{Tulin:2017ara}
  S.~Tulin and H.~B.~Yu,
  arXiv:1705.02358 [hep-ph].

\bibitem{Clowe:2003tk}
  D.~Clowe, A.~Gonzalez and M.~Markevitch,
  Astrophys.\ J.\  {\bf 604}, 596 (2004)
  doi:10.1086/381970
  [astro-ph/0312273].

\bibitem{Markevitch:2003at}
  M.~Markevitch {\it et al.},
  Astrophys.\ J.\  {\bf 606}, 819 (2004)
  doi:10.1086/383178
  [astro-ph/0309303].

\bibitem{Randall:2007ph}
  S.~W.~Randall, M.~Markevitch, D.~Clowe, A.~H.~Gonzalez and M.~Bradac,
  Astrophys.\ J.\  {\bf 679}, 1173 (2008)
  doi:10.1086/587859
  [arXiv:0704.0261 [astro-ph]].

\bibitem{Kahlhoefer:2013dca}
  F.~Kahlhoefer, K.~Schmidt-Hoberg, M.~T.~Frandsen and S.~Sarkar,
  Mon.\ Not.\ Roy.\ Astron.\ Soc.\  {\bf 437}, no. 3, 2865 (2014)
  doi:10.1093/mnras/stt2097
  [arXiv:1308.3419 [astro-ph.CO]].

\bibitem{Harvey:2015hha}
  D.~Harvey, R.~Massey, T.~Kitching, A.~Taylor and E.~Tittley,
  Science {\bf 347}, 1462 (2015)
  doi:10.1126/science.1261381
  [arXiv:1503.07675 [astro-ph.CO]].

\bibitem{Wittman:2017gxn}
  D.~Wittman, N.~Golovich and W.~A.~Dawson,
  arXiv:1701.05877 [astro-ph.CO].


\bibitem{Ackerman:mha}
  L.~Ackerman, M.~R.~Buckley, S.~M.~Carroll and M.~Kamionkowski,
  Phys.\ Rev.\ D {\bf 79}, 023519 (2009)
  doi:10.1103/PhysRevD.79.023519, 10.1142/9789814293792\_0021
  [arXiv:0810.5126 [hep-ph]].

\bibitem{Buckley:2009in}
  M.~R.~Buckley and P.~J.~Fox,
  Phys.\ Rev.\ D {\bf 81}, 083522 (2010)
  doi:10.1103/PhysRevD.81.083522
  [arXiv:0911.3898 [hep-ph]].

\bibitem{Loeb:2010gj}
  A.~Loeb and N.~Weiner,
  Phys.\ Rev.\ Lett.\  {\bf 106}, 171302 (2011)
  doi:10.1103/PhysRevLett.106.171302
  [arXiv:1011.6374 [astro-ph.CO]].



\bibitem{Feng:2009hw}
  J.~L.~Feng, M.~Kaplinghat and H.~B.~Yu,
  Phys.\ Rev.\ Lett.\  {\bf 104}, 151301 (2010)
  doi:10.1103/PhysRevLett.104.151301
  [arXiv:0911.0422 [hep-ph]].

\bibitem{Tulin:2012wi}
  S.~Tulin, H.~B.~Yu and K.~M.~Zurek,
  Phys.\ Rev.\ Lett.\  {\bf 110}, no. 11, 111301 (2013)
  doi:10.1103/PhysRevLett.110.111301
  [arXiv:1210.0900 [hep-ph]].

\bibitem{Tulin:2013teo}
  S.~Tulin, H.~B.~Yu and K.~M.~Zurek,
  Phys.\ Rev.\ D {\bf 87}, no. 11, 115007 (2013)
  doi:10.1103/PhysRevD.87.115007
  [arXiv:1302.3898 [hep-ph]].

\bibitem{Aarssen:2012fx}
  L.~G.~van den Aarssen, T.~Bringmann and C.~Pfrommer,
  Phys.\ Rev.\ Lett.\  {\bf 109}, 231301 (2012)
  doi:10.1103/PhysRevLett.109.231301
  [arXiv:1205.5809 [astro-ph.CO]].

\bibitem{Cyr-Racine:2015ihg}
  F.~Y.~Cyr-Racine, K.~Sigurdson, J.~Zavala, T.~Bringmann, M.~Vogelsberger and C.~Pfrommer,
  Phys.\ Rev.\ D {\bf 93}, no. 12, 123527 (2016)
  doi:10.1103/PhysRevD.93.123527
  [arXiv:1512.05344 [astro-ph.CO]].

\bibitem{Nozzoli:2016coi}
  F.~Nozzoli,
  Astropart.\ Phys.\  {\bf 91}, 22 (2017)
  doi:10.1016/j.astropartphys.2017.03.005
  [arXiv:1608.00405 [astro-ph.IM]].

\bibitem{Foot:2014uba} 
  R.~Foot and S.~Vagnozzi,
  Phys.\ Rev.\ D {\bf 91}, 023512 (2015)
  doi:10.1103/PhysRevD.91.023512
  [arXiv:1409.7174 [hep-ph]].

\bibitem{Bringmann:2016din}
  T.~Bringmann, F.~Kahlhoefer, K.~Schmidt-Hoberg and P.~Walia,
  Phys.\ Rev.\ Lett.\  {\bf 118}, no. 14, 141802 (2017)
  doi:10.1103/PhysRevLett.118.141802
  [arXiv:1612.00845 [hep-ph]].

\bibitem{Kahlhoefer:2017umn}
  F.~Kahlhoefer, K.~Schmidt-Hoberg and S.~Wild,
  JCAP08(2017)003
  [arXiv:1704.02149 [hep-ph]].



\bibitem{Pospelov:2007mp}
  M.~Pospelov, A.~Ritz and M.~B.~Voloshin,
  Phys.\ Lett.\ B {\bf 662}, 53 (2008)
  doi:10.1016/j.physletb.2008.02.052
  [arXiv:0711.4866 [hep-ph]].


\bibitem{McDonald:2001vt}
  J.~McDonald,
  Phys.\ Rev.\ Lett.\  {\bf 88}, 091304 (2002)
  doi:10.1103/PhysRevLett.88.091304
  [hep-ph/0106249].

\bibitem{Hall:2009bx}
  L.~J.~Hall, K.~Jedamzik, J.~March-Russell and S.~M.~West,
  JHEP {\bf 1003}, 080 (2010)
  doi:10.1007/JHEP03(2010)080
  [arXiv:0911.1120 [hep-ph]].

\bibitem{Bernal:2017kxu}
  N.~Bernal, M.~Heikinheimo, T.~Tenkanen, K.~Tuominen and V.~Vaskonen,
  arXiv:1706.07442 [hep-ph].

\bibitem{Cheung:2010gj}
  C.~Cheung, G.~Elor, L.~J.~Hall and P.~Kumar,
  JHEP {\bf 1103}, 042 (2011)
  doi:10.1007/JHEP03(2011)042
  [arXiv:1010.0022 [hep-ph]].
  
\bibitem{Chu:2011be}
  X.~Chu, T.~Hambye and M.~H.~G.~Tytgat,
  JCAP {\bf 1205}, 034 (2012)
  doi:10.1088/1475-7516/2012/05/034
  [arXiv:1112.0493 [hep-ph]].

\bibitem{Campbell:2015fra} 
  R.~Campbell, S.~Godfrey, H.~E.~Logan, A.~D.~Peterson and A.~Poulin,
  Phys.\ Rev.\ D {\bf 92}, no. 5, 055031 (2015)
  doi:10.1103/PhysRevD.92.055031
  [arXiv:1505.01793 [hep-ph]].
  
\bibitem{Kang:2015aqa} 
  Z.~Kang,
  Phys.\ Lett.\ B {\bf 751}, 201 (2015)
  doi:10.1016/j.physletb.2015.10.031
  [arXiv:1505.06554 [hep-ph]].
  
\bibitem{Bernal:2015ova}
  N.~Bernal, X.~Chu, C.~Garcia-Cely, T.~Hambye and B.~Zaldivar,
  JCAP {\bf 1603}, no. 03, 018 (2016)
  doi:10.1088/1475-7516/2016/03/018
  [arXiv:1510.08063 [hep-ph]].
  
\bibitem{Bernal:2015xba} 
  N.~Bernal and X.~Chu,
  JCAP {\bf 1601}, 006 (2016)
  doi:10.1088/1475-7516/2016/01/006
  [arXiv:1510.08527 [hep-ph]].

  
\bibitem{Ayazi:2015jij} 
  S.~Yaser Ayazi, S.~M.~Firouzabadi and S.~P.~Zakeri,
  J.\ Phys.\ G {\bf 43}, no. 9, 095006 (2016)
  doi:10.1088/0954-3899/43/9/095006
  [arXiv:1511.07736 [hep-ph]].
  
\bibitem{Bernal:2017mqb} 
  N.~Bernal, X.~Chu and J.~Pradler,
  Phys.\ Rev.\ D {\bf 95}, no. 11, 115023 (2017)
  doi:10.1103/PhysRevD.95.115023
  [arXiv:1702.04906 [hep-ph]].




\bibitem{Hambye:2008bq}
  T.~Hambye,
  JHEP {\bf 0901}, 028 (2009)
  doi:10.1088/1126-6708/2009/01/028
  [arXiv:0811.0172 [hep-ph]].

\bibitem{Lebedev:2011iq}
  O.~Lebedev, H.~M.~Lee and Y.~Mambrini,
  Phys.\ Lett.\ B {\bf 707}, 570 (2012)
  doi:10.1016/j.physletb.2012.01.029
  [arXiv:1111.4482 [hep-ph]].

\bibitem{Farzan:2012hh}
  Y.~Farzan and A.~R.~Akbarieh,
  JCAP {\bf 1210}, 026 (2012)
  doi:10.1088/1475-7516/2012/10/026
  [arXiv:1207.4272 [hep-ph]].

\bibitem{Baek:2012se}
  S.~Baek, P.~Ko, W.~I.~Park and E.~Senaha,
  JHEP {\bf 1305}, 036 (2013)
  doi:10.1007/JHEP05(2013)036
  [arXiv:1212.2131 [hep-ph]].

\bibitem{Baek:2014jga}
  S.~Baek, P.~Ko and W.~I.~Park,
  Phys.\ Rev.\ D {\bf 90}, no. 5, 055014 (2014)
  doi:10.1103/PhysRevD.90.055014
  [arXiv:1405.3530 [hep-ph]].


\bibitem{Duch:2015jta}
  M.~Duch, B.~Grzadkowski and M.~McGarrie,
  JHEP {\bf 1509}, 162 (2015)
  doi:10.1007/JHEP09(2015)162
  [arXiv:1506.08805 [hep-ph]].

\bibitem{Duch:2015cxa}
  M.~Duch, B.~Grzadkowski and M.~McGarrie,
  Acta Phys.\ Polon.\ B {\bf 46}, no. 11, 2199 (2015)
  doi:10.5506/APhysPolB.46.2199
  [arXiv:1510.03413 [hep-ph]].

\bibitem{Karam:2015jta}
  A.~Karam and K.~Tamvakis,
  Phys.\ Rev.\ D {\bf 92}, no. 7, 075010 (2015)
  doi:10.1103/PhysRevD.92.075010
  [arXiv:1508.03031 [hep-ph]].
\bibitem{Karam:2016rsz}
  A.~Karam and K.~Tamvakis,
  Phys.\ Rev.\ D {\bf 94}, no. 5, 055004 (2016)
  doi:10.1103/PhysRevD.94.055004
  [arXiv:1607.01001 [hep-ph]].
\bibitem{Arcadi:2016kmk}
  G.~Arcadi, C.~Gross, O.~Lebedev, Y.~Mambrini, S.~Pokorski and T.~Toma,
  JHEP {\bf 1612}, 081 (2016)
  doi:10.1007/JHEP12(2016)081
  [arXiv:1611.00365 [hep-ph]].
\bibitem{Heikinheimo:2017ofk}
  M.~Heikinheimo, T.~Tenkanen and K.~Tuominen,
  Phys.\ Rev.\ D {\bf 96}, no. 2, 023001 (2017)
  doi:10.1103/PhysRevD.96.023001
  [arXiv:1704.05359 [hep-ph]].




\bibitem{Quiros:1999jp}
  M.~Quiros,
  hep-ph/9901312.

\bibitem{Katz:2014bha}
  A.~Katz and M.~Perelstein,
  JHEP {\bf 1407}, 108 (2014)
  doi:10.1007/JHEP07(2014)108
  [arXiv:1401.1827 [hep-ph]].

\bibitem{LanHEP1}
  A.~V.~Semenov,
  hep-ph/9608488, hep-ph/0208011.

\bibitem{LanHEP3}
  A.~Semenov,
  Comput.\ Phys.\ Commun.\  {\bf 180}, 431 (2009)
  doi:10.1016/j.cpc.2008.10.012
  [arXiv:0805.0555 [hep-ph]].

\bibitem{CalcHEP}
  A.~Belyaev, N.~D.~Christensen and A.~Pukhov,
  Comput.\ Phys.\ Commun.\  {\bf 184}, 1729 (2013)
  doi:10.1016/j.cpc.2013.01.014
  [arXiv:1207.6082 [hep-ph]].

\bibitem{Khrapak:2003kjw}
  S.~A.~Khrapak, A.~V.~Ivlev, G.~E.~Morfill and S.~K.~Zhdanov,
  Phys.\ Rev.\ Lett.\  {\bf 90}, no. 22, 225002 (2003).
  doi:10.1103/PhysRevLett.90.225002



\bibitem{Cline:2013gha}
  J.~M.~Cline, K.~Kainulainen, P.~Scott and C.~Weniger,
  Phys.\ Rev.\ D {\bf 88}, 055025 (2013)
  Erratum: [Phys.\ Rev.\ D {\bf 92}, no. 3, 039906 (2015)]
  doi:10.1103/PhysRevD.92.039906, 10.1103/PhysRevD.88.055025
  [arXiv:1306.4710 [hep-ph]].
  
\bibitem{Alarcon:2011zs} 
  J.~M.~Alarcon, J.~Martin Camalich and J.~A.~Oller,
  Phys.\ Rev.\ D {\bf 85}, 051503 (2012)
  doi:10.1103/PhysRevD.85.051503
  [arXiv:1110.3797 [hep-ph]].
  
\bibitem{Alarcon:2012nr} 
  J.~M.~Alarcon, L.~S.~Geng, J.~Martin Camalich and J.~A.~Oller,
  Phys.\ Lett.\ B {\bf 730}, 342 (2014)
  doi:10.1016/j.physletb.2014.01.065
  [arXiv:1209.2870 [hep-ph]].

\bibitem{Akerib:2016vxi}
  D.~S.~Akerib {\it et al.} [LUX Collaboration],
  Phys.\ Rev.\ Lett.\  {\bf 118}, no. 2, 021303 (2017)
  doi:10.1103/PhysRevLett.118.021303
  [arXiv:1608.07648 [astro-ph.CO]].

\bibitem{PandaX16}
  A.~Tan {\it et al.} [PandaX-II Collaboration],
  Phys.\ Rev.\ Lett.\  {\bf 117}, no. 12, 121303 (2016)
  doi:10.1103/PhysRevLett.117.121303
  [arXiv:1607.07400 [hep-ex]].

\bibitem{XENON1T}
  E.~Aprile {\it et al.} [XENON Collaboration],
  arXiv:1705.06655 [astro-ph.CO].



\bibitem{Geng:2016uqt}
  C.~Q.~Geng, D.~Huang, C.~H.~Lee and Q.~Wang,
  JCAP {\bf 1608}, no. 08, 009 (2016)
  doi:10.1088/1475-7516/2016/08/009
  [arXiv:1605.05098 [hep-ph]].


\bibitem{Geng:2017ypy}
  C.~Q.~Geng, D.~Huang and C.~H.~Lee,
  arXiv:1705.06546 [hep-ph].



\bibitem{Jedamzik:2009uy}
  K.~Jedamzik and M.~Pospelov,
  New J.\ Phys.\  {\bf 11}, 105028 (2009)
  doi:10.1088/1367-2630/11/10/105028
  [arXiv:0906.2087 [hep-ph]].

\bibitem{Kawasaki:2004qu}
  M.~Kawasaki, K.~Kohri and T.~Moroi,
  Phys.\ Rev.\ D {\bf 71}, 083502 (2005)
  doi:10.1103/PhysRevD.71.083502
  [astro-ph/0408426].

\bibitem{Kawasaki:2017bqm}
  M.~Kawasaki, K.~Kohri, T.~Moroi and Y.~Takaesu,
  arXiv:1709.01211 [hep-ph].

\bibitem{Berger:2016vxi}
  J.~Berger, K.~Jedamzik and D.~G.~E.~Walker,
  JCAP {\bf 1611}, 032 (2016)
  doi:10.1088/1475-7516/2016/11/032
  [arXiv:1605.07195 [hep-ph]].

\bibitem{Ade:2015xua}
  P.~A.~R.~Ade {\it et al.} [Planck Collaboration],
  Astron.\ Astrophys.\  {\bf 594}, A13 (2016)
  doi:10.1051/0004-6361/201525830
  [arXiv:1502.01589 [astro-ph.CO]].

\bibitem{Padmanabhan:2005es}
  N.~Padmanabhan and D.~P.~Finkbeiner,
  Phys.\ Rev.\ D {\bf 72}, 023508 (2005)
  doi:10.1103/PhysRevD.72.023508
  [astro-ph/0503486].

\bibitem{Slatyer:2015jla}
  T.~R.~Slatyer,
  Phys.\ Rev.\ D {\bf 93}, no. 2, 023527 (2016)
  doi:10.1103/PhysRevD.93.023527
  [arXiv:1506.03811 [hep-ph]].

\bibitem{Slatyer:2015kla}
  T.~R.~Slatyer,
  Phys.\ Rev.\ D {\bf 93}, no. 2, 023521 (2016)
  doi:10.1103/PhysRevD.93.023521
  [arXiv:1506.03812 [astro-ph.CO]].


\bibitem{ArkaniHamed:2008qn}
  N.~Arkani-Hamed, D.~P.~Finkbeiner, T.~R.~Slatyer and N.~Weiner,
  Phys.\ Rev.\ D {\bf 79}, 015014 (2009)
  doi:10.1103/PhysRevD.79.015014
  [arXiv:0810.0713 [hep-ph]].

\bibitem{Sommerfeld}
  A.~Sommerfeld, Ann.\ Phys.\ (Berlin) {\bf 403}, 257 (1931).

\bibitem{Cassel:2009wt}
  S.~Cassel,
  J.\ Phys.\ G {\bf 37}, 105009 (2010)
  doi:10.1088/0954-3899/37/10/105009
  [arXiv:0903.5307 [hep-ph]].

\bibitem{Iengo:2009xf}
  R.~Iengo,
  arXiv:0903.0317 [hep-ph].

\bibitem{Slatyer:2009vg}
  T.~R.~Slatyer,
  JCAP {\bf 1002}, 028 (2010)
  doi:10.1088/1475-7516/2010/02/028
  [arXiv:0910.5713 [hep-ph]].

\bibitem{Elor:2015bho}
  G.~Elor, N.~L.~Rodd, T.~R.~Slatyer and W.~Xue,
  JCAP {\bf 1606}, no. 06, 024 (2016)
  doi:10.1088/1475-7516/2016/06/024
  [arXiv:1511.08787 [hep-ph]].

\bibitem{Bergstrom:2013jra}
  L.~Bergstrom, T.~Bringmann, I.~Cholis, D.~Hooper and C.~Weniger,
  Phys.\ Rev.\ Lett.\  {\bf 111}, 171101 (2013)
  doi:10.1103/PhysRevLett.111.171101
  [arXiv:1306.3983 [astro-ph.HE]].

\bibitem{Hooper:2012gq}
  D.~Hooper and W.~Xue,
  Phys.\ Rev.\ Lett.\  {\bf 110}, no. 4, 041302 (2013)
  doi:10.1103/PhysRevLett.110.041302
  [arXiv:1210.1220 [astro-ph.HE]].

\bibitem{Ibarra:2013zia}
  A.~Ibarra, A.~S.~Lamperstorfer and J.~Silk,
  Phys.\ Rev.\ D {\bf 89}, no. 6, 063539 (2014)
  doi:10.1103/PhysRevD.89.063539
  [arXiv:1309.2570 [hep-ph]].

\bibitem{AMS1}
  M.~Aguilar {\it et al.} [AMS Collaboration],
  Phys.\ Rev.\ Lett.\  {\bf 113}, 121102 (2014).
  doi:10.1103/PhysRevLett.113.121102

\bibitem{AMS2}
  L.~Accardo {\it et al.} [AMS Collaboration],
  Phys.\ Rev.\ Lett.\  {\bf 113}, 121101 (2014).
  doi:10.1103/PhysRevLett.113.121101

\bibitem{Ackermann:2015zua}
  M.~Ackermann {\it et al.} [Fermi-LAT Collaboration],
  Phys.\ Rev.\ Lett.\  {\bf 115}, no. 23, 231301 (2015)
  doi:10.1103/PhysRevLett.115.231301
  [arXiv:1503.02641 [astro-ph.HE]].

\bibitem{Adams:1998nr}
  J.~A.~Adams, S.~Sarkar and D.~W.~Sciama,
  Mon.\ Not.\ Roy.\ Astron.\ Soc.\  {\bf 301}, 210 (1998)
  doi:10.1046/j.1365-8711.1998.02017.x
  [astro-ph/9805108].

\bibitem{Chen:2003gz}
  X.~L.~Chen and M.~Kamionkowski,
  Phys.\ Rev.\ D {\bf 70}, 043502 (2004)
  doi:10.1103/PhysRevD.70.043502
  [astro-ph/0310473].

\bibitem{Slatyer:2016qyl}
  T.~R.~Slatyer and C.~L.~Wu,
  Phys.\ Rev.\ D {\bf 95}, no. 2, 023010 (2017)
  doi:10.1103/PhysRevD.95.023010
  [arXiv:1610.06933 [astro-ph.CO]].

\bibitem{Essig:2013goa}
  R.~Essig, E.~Kuflik, S.~D.~McDermott, T.~Volansky and K.~M.~Zurek,
  JHEP {\bf 1311}, 193 (2013)
  doi:10.1007/JHEP11(2013)193
  [arXiv:1309.4091 [hep-ph]].

\bibitem{Boddy:2015efa}
  K.~K.~Boddy and J.~Kumar,
  Phys.\ Rev.\ D {\bf 92}, no. 2, 023533 (2015)
  doi:10.1103/PhysRevD.92.023533
  [arXiv:1504.04024 [astro-ph.CO]].

\bibitem{Riemer-Sorensen:2015kqa}
  S.~Riemer-Sørensen {\it et al.},
  Astrophys.\ J.\  {\bf 810}, no. 1, 48 (2015)
  doi:10.1088/0004-637X/810/1/48
  [arXiv:1507.01378 [astro-ph.CO]].

\bibitem{Dittmaier:2011ti}
  S.~Dittmaier {\it et al.} [LHC Higgs Cross Section Working Group],
  doi:10.5170/CERN-2011-002
  arXiv:1101.0593 [hep-ph].

\bibitem{Heinemeyer:2013tqa}
  S.~Heinemeyer {\it et al.} [LHC Higgs Cross Section Working Group],
  doi:10.5170/CERN-2013-004
  arXiv:1307.1347 [hep-ph].

\bibitem{Edsjo:1997bg}
  J.~Edsjo and P.~Gondolo,
  Phys.\ Rev.\ D {\bf 56}, 1879 (1997)
  doi:10.1103/PhysRevD.56.1879
  [hep-ph/9704361].

\end{thebibliography}
\end{document}